\documentclass[epj]{svjour}
\journalname{EPJB}
\usepackage{amsmath}
\usepackage{amssymb}
\usepackage{epsf}
\newlength{\figwidth}

\begin{document}

\title{Delocalization effects and charge reorganizations induced by
       repulsive interactions in strongly disordered chains}
\titlerunning{Delocalization due to repulsive interactions in strongly
       disordered chains}
\author{Dietmar~Weinmann\inst{1,2} \and Peter~Schmitteckert\inst{1,3}
        \and Rodolfo~A.~Jalabert\inst{1} \and Jean-Louis~Pichard\inst{4}}

\institute{Institut de Physique et Chimie des Mat\'eriaux de Strasbourg,
           UMR 7504, CNRS-ULP\\ 23 rue du Loess, 67037 Strasbourg
           Cedex, France 
           \and 
           Institut f\"ur Physik, Universit\"at Augsburg,
           86135 Augsburg, Germany 
           \and 
           Fridolinstr.\ 19, 68753 Wagh\"ausel, Germany 
           \and
           CEA, Service de Physique de l'Etat Condens\'e,
           Centre d'Etudes de Saclay, 91191 Gif-sur-Yvette Cedex, France}

\date{November 6, 2000} 

\abstract{
 We study the delocalization effect of a short-range repulsive
interaction on the ground state of a finite density of spinless
fermions in strongly disordered one dimensional lattices. The density 
matrix renormalization group method is used to explore the charge density 
and the sensitivity of the ground state energy with respect to the 
boundary condition (the persistent current) for a wide range of
parameters (carrier density, interaction and disorder). Analytical 
approaches are developed and allow to understand some mechanisms and
limiting conditions. For weak interaction strength, 
one has a Fermi glass of Anderson localized states, while in the
opposite limit of strong interaction, one has a correlated 
array of charges (Mott insulator). In the two cases, the system is 
strongly insulating and the ground state energy is essentially invariant 
under a twist of the boundary conditions. Reducing the interaction 
strength from large to intermediate values, the quantum melting of the 
solid array gives rise to a more homogeneous distribution of charges, 
and the ground state energy changes when the boundary conditions are twisted. 
In individual chains, this melting occurs by abrupt steps located at
sample-dependent values of the interaction where an (avoided) level 
crossing between the ground state and the first excitation can be 
observed. Important charge reorganizations take place at the 
avoided crossings and the persistent currents are strongly enhanced
around the corresponding interaction value. 
These large delocalization effects become smeared and reduced after 
ensemble averaging. They mainly characterize half filling and strong 
disorder, but they persist away of this optimal condition.
\PACS{ 
       {72.15.-v}{Electronic conduction in metals and alloy} \and
       {73.20.Dx}{Electron states in low-dimensional structures} \and 
       {72.10.Bg}{General formulation of transport theory} \and
       {05.60.Gg}{Quantum transport}
      }
} 
\maketitle
 

\section{Introduction}
\label{sec:intro}

One of the beauties of Condensed Matter Physics is the
possibility of understanding a variety of phenomena within a
(weakly interacting) quasi-particle approach, despite the always present 
strong Coulomb interaction between electrons. As it has been understood 
from the early days \cite{A&M}, the applicability of one-particle 
approaches can be traced to the Pauli exclusion principle, and in first
approximation the interactions simply account for a renormalization
of single-particle quantities (like the effective mass or the mean
field potential felt by individual electrons).

 This traditional view is challenged when studying artificially 
confined mesoscopic systems or very dilute low-dimensional electron gases. 
Mesoscopic phenomena have their origin in the coherence of electronic wave
functions across a small sample. The reduced dimensions are expected to 
render electronic correlations more important than in the bulk.
Going to lower dimensions and/or very dilute limits results in a
poorer screening of the electron-electron interaction, enhancing the
role of Coulomb repulsions. 
When the disorder is strong, Anderson localization becomes also detrimental 
to screening, thereby further magnifying the role of interactions.

 The above considerations are closely related to three experimental 
findings which have recently dominated the attention in Mesoscopic Physics.
Firstly, the large values of the persistent current measured in metallic 
mesoscopic rings \cite{levy90,Webb1,Webb2} 
cannot be accounted by the theoretical predictions based on the
single electron picture \cite{ensemble}, nor by perturbative approaches 
taking into account (to infinite order) the effect of electron-electron 
interactions \cite{Eckern,ESra}. Secondly, the discovery of a metallic 
behavior in two-dimensional gases of electrons (Si-MOSFET \cite{Kravchenko}) 
or holes (GaAs heterostructures \cite{Pepper} and SiGe quantum 
wells \cite{SiGe}) is at odds with the conclusions of non-interacting 
theories that predict an insulating behavior for any value of the 
disorder strength \cite{Go4}. For a recent review on the
metal-insulator transition in $2D$ electron and hole gases see 
Ref.~\cite{abrahams}. The metallic behavior is observed in a
parameter region of the system which is believed to be limited by its 
crystallization threshold \cite{benenti_new,spivak}. Finally, the 
conventional view of an infinite zero-temperature quasiparticle
lifetime at the Fermi energy has been questioned by the interpretation
of measurements yielding a saturation of the electronic decoherence rate with
decreasing temperature \cite{Moha}. While a complete understanding of 
these findings is still missing, it is widely accepted that they are 
influenced in a non-trivial way by large interaction effects. Recent 
theoretical attempts suggest a possible relation 
between these different phenomena \cite{Kravtsov,Moha2,Schwab}.

In this work we study the persistent current and the localization within
a model of spinless electrons on a disordered one-dimensional ($1D$) 
ring with nearest-neighbor interactions. Clearly, we do not aim to 
describe the metallic quasi-one dimensional rings of 
Refs.~\cite{levy90,Webb1,Webb2} nor the two-dimensional 
electron gas where the metal-insulator transition has been observed.
However, the interplay between localization and interaction can be
readily studied in this simple model. The interest on disordered
one-dimensional interacting models of fermions (with and without
spin) can also be assessed from the large variety of analytical
\cite{Luther,gs2,Shankar,SetS,GetS,AetW} and numerical 
\cite{pang,AetB,BPM,Kato,tsiper,peter,Jeon,Chiappe} techniques that 
have been applied to them.

In the absence of interactions, a $1D$ disordered system is an
Anderson insulator, with the electron wave-functions localized
on the scale of the one-particle localization length $\xi_1$. 
The dependence of the conductance on the size $M$ of the system is
given by an exponential decrease, the characteristic length scale
being determined by $\xi_1$.
If the system is closed into a ring threaded by a magnetic flux, its
orbital response (the persistent current) 
also scales exponentially with $M/\xi_1$ \cite{Cheung}.

In rotational invariant continuum systems the persistent current does 
not depend on any kind of interactions \cite{SetS,AetW}. This is directly 
connected to the fact that the Drude weight (to be defined in the sequel)
is not affected by electron-electron interactions in Galilean
invariant systems \cite{Okabe}. 
When the rotational invariance is broken by a lattice or by the presence of 
disorder, interactions can modify the value of the persistent current 
\cite{peter}.

The problem of spinless fermions in a disorder-free chain
with nearest-neighbor interactions is exactly solvable \cite{Luther}.
In particular, at half filling we have a Mott insulator with a
finite gap and a charge density wave concentrated on alternating
sites of the lattice. Attractive interactions favor an inhomogeneous 
density (clustering). Repulsive interactions favor a homogeneous density 
(charge density wave or Mott insulator). Disorder tends to distort those 
arrangements by favoring the occupancy of the low potential sites. 
In the intermediate regime between the Anderson
and Mott insulators, disorder, interaction and kinetic energy 
are relevant. The competition between disorder and interactions 
that we study throughout our work exhibits then a non-trivial character. 

The localized character of an electron system determines the
behavior of the conductance, which is a transport property, as well 
as the persistent current, which is a thermodynamic property. As first
shown by Kohn \cite{kohn}, in the zero temperature limit, both
properties can be related. We recall the basic ingredients of
such a relationship \cite{SetS,kohn,scala} for the particular case of
the insulating regime. 

The linear response of the electronic system to a spatially uniform,
time-dependent electric field is the frequency-dependent conductivity
\begin{equation}\label{eq:conductivity}
\sigma(\omega)=\sigma_1(\omega)+{\rm i} \sigma_2(\omega)\, .
\end{equation}

A $1D$ ring containing $M$ sites, threaded by a magnetic flux
$\Phi$ has a flux-dependent
many-particle ground state energy $E(\Phi)$. We choose units such that
$\hbar=e^2=c=a=1$ ($a$ is the lattice constant), and
$\Phi=2\pi$ corresponds to one flux quantum threading the ring.
Kohn showed that the second derivative of $E(\Phi)$ (the charge
stiffness or Kohn curvature) 
\begin{equation}
D_c = M \left.\left(\frac{{\rm d}^2 E(\Phi)}{{\rm d} \Phi^2}\right)
\right|_{\Phi=0} 
\end{equation}
is related to the imaginary part $\sigma_2(\omega)$ of the conductivity 
through
\begin{equation}\label{eq:drude-im}
D_c = \lim_{\omega\rightarrow 0} \omega \, \sigma_2(\omega) \, .
\end{equation}

Using the Kramers-Kronig relations between the real and the imaginary
part of the conductivity (see, e.g.\ \cite{SetS}), it is found that
$D_c$ also gives the weight of the zero-frequency peak in the real
part of the conductivity
\begin{equation}
\sigma_1(\omega) = \pi D_c \delta(\omega) + \sigma_1^{\rm reg}(\omega) 
\, .
\end{equation} 
Therefore, $D_c$ is sometimes called the Drude weight. This establishes a link
between transport properties and persistent currents. 

In the insulating regime the amplitude of the flux-dependent oscillation
of the ground state energy is typically much smaller than the energy gap 
between the many-body ground state and the first excited state. Therefore, 
it is easy to see ({\em i.e.} Sec.~\ref{sec:phymec}) that a perturbation
theory in the hopping matrix elements across the boundary is enough to
describe the flux dependence of the ground state energy, which yields
\begin{equation}
E(\Phi)=E(0)-\frac{\Delta E}{2}(1-\cos\Phi) \, .
\label{eq:Phidep}
\end{equation}
Here $\Delta E = E(0)\!-\!E(\pi)$ can also be interpreted as the
difference of ground state energy between periodic ($\Phi\!=\!0$) and 
anti-periodic ($\Phi\!=\!\pi$) boundary conditions, since a magnetic
flux through the ring is equivalent to introducing a change of the
boundary conditions. The sign of $\Delta E$ depends only on the parity
of the number of particles $N$ ($\Delta E < 0$ for odd $N$ and 
$\Delta E > 0$ for even $N$) \cite{Leggett,loss}.

The simple $\Phi$-dependence of the ground state energy in 
Eq.~(\ref{eq:Phidep}) allows to relate $\Delta E$ to the persistent
current
\begin{equation}
J(\Phi)= - \frac{{\rm d} E(\Phi)}{{\rm d} \Phi} = \frac{\Delta E}{2}
\sin{\Phi} \, ,
\end{equation}
and the Drude weight
\begin{equation}
D_{c} = - \frac{M}{2} \Delta E \, .
\end{equation}

In this paper we will work extensively with the {\em phase sensitivity}
\begin{equation}\label{eq:defphasen}
D = (-1)^N \frac{M}{2} \Delta E \, ,
\end{equation}
that for a {\em strongly disordered one-dimensional ring} is simply
given by the absolute value of the Drude weight and, at the same time, 
determines the magnitude of the persistent current.

The possibility of a negative charge stiffness (or $D_c$) arising for
spinless fermions and Hubbard rings \cite{Staff91,Fye91} indicates that
the orbital response may be paramagnetic. This is a peculiar behavior
since $D_c$ is believed to determine the zero-frequency behavior of 
$\sigma_1$. 

The previous numerical work on interacting disordered rings has necessarily
been restricted to finite samples \cite{AetB,BPM} or it has relied
on strong approximations (like Hartree-Fock \cite{Kato,Jeon}). 
Direct diagonalization of small systems (lattices with $M\!=\!6$
and $10$ sites at half filling) with Coulomb interaction \cite{AetB}
yielded an (impurity) average persistent current that is suppressed
by effects of the interactions except for strong disorder and
weak interaction strength, where it is weakly enhanced. Direct
diagonalization in one-dimensional rings of spinless fermions with
short-range interactions in lattices of up to 20 sites \cite{BPM}
found that both, disorder and interactions, always decrease the
persistent current by localizing the electrons. The above simulation
deals with values of the disorder that are not strong in comparison 
with the typical kinetic energy of the electrons. Similar results
have been obtained in this regime using the density matrix renormalization
group (DMRG) method \cite{peter}. Hartree-Fock calculations also 
yielded a suppression of the persistent current as the strength of the 
interaction increases \cite{Kato}.

The conclusions that we extract from our numerical computations of
the phase sensitivity are somehow different than that of the previous
numerical studies. Working at large disorder, we find that when 
increasing the strength of the interactions, abrupt charge reorganizations 
of the many-body ground state take place (at sample-dependent values 
of the interaction strength) and are associated with anomalously large 
persistent currents \cite{sjwp,wpsj,PhysicaE}. In this work we extend 
our previous numerical calculations showing such an effect and we also 
present analytical work aiding towards its understanding. We point out
to the importance of considering the physics of individual samples and
we show that the delocalization effect of interactions persists in 
the thermodynamic limit.

A sizeable increase of the persistent current with the strength of the
interaction had been obtained for moderately disordered $1D$ electronic systems
{\em with spin} (Anderson-Hubbard model) from renormalization group
approaches \cite{GetS} or perturbation and numerical calculations
\cite{Chiappe}. Our results show that the spin does not seem to be
a necessary ingredient to obtain an enhancement of persistent currents due
to interactions. Even without spin, repulsive interactions can increase 
the persistent current, provided the disorder is important enough. This lets 
us expect an even more dramatic increase at strong disorder in models with 
spin.

It is interesting to remark that the enhancement of transport properties
by the effect of a repulsive interaction has been proposed in other
contexts than the one of this work. A system with strong binary disorder, 
where the two possible values of the disorder give rise to two separated 
bands, presents in the absence of interactions a gap if the filling is 
such that the lower band is filled completely \cite{ulmke}.
Then, the broadening of the bands due to the interaction can lead to an 
overlap allowing for metallic behavior. Also, in a system of 
two interacting particles on a disordered chain it has been shown
that the localization length is enhanced by the effect of interactions
\cite{shepelyansky,tip2,tip3,tip4}.

The remainder of the paper is organized as follows.
In Section \ref{sec:modmet}, we present the model and the numerical method.
In Section \ref{sec:halff}, we perform numerical studies for the ground state 
structure and the phase sensitivity for the case of half filling.
A scaling with the system size allows to demonstrate the
delocalization effect of repulsive interactions in the presence of
strong disorder.
In Section \ref{sec:phymec}, analytical work is presented, which aims
to describe the basic physical mechanisms 
leading to the effects which were found numerically. In Section 
\ref{sec:deponfil}, we extend the numerical studies away from half
filling and establish the criterium for observing a delocalization
effect due to the interactions. Finally, we present in Section
\ref{sec:conclusion} our conclusions and outline some open problems.

\section{Model and Method} \label{sec:modmet}
\subsection{Model Hamiltonian}
 We consider spinless fermions on a chain with nearest-neighbor interaction
\begin{equation}\label{hamiltonian}
H=-t \sum_{i=1}^{M} (c_i^{\dagger} c_{i-1} + c^{\dagger}_{i-1}c_i) 
+\sum_{i=1}^{M} v_i n_i + U \sum_{i=1}^{M} n_i n_{i-1}
\end{equation}
and twisted boundary conditions, $c_0=\exp({\rm i}\Phi) c_M$. The operators 
$c_i$  ($c^{\dagger}_{i}$) destroy (create) a particle on site $i$ and 
$n_i=c^{\dagger}_ic_i$ is the occupation operator. The on-site random 
energies $v_i$ are drawn from a box distribution of width $W$. The strength 
of the disorder $W$ and the interaction $U$ are measured in units of the 
kinetic energy scale ( $t\!=\!1$). The use of twisted boundary conditions 
allows to represent a ring of $M$ sites pierced by a flux $\Phi$.

\subsection{Numerical Method}
 
 The numerical results are obtained with the density matrix renormalization
group (DMRG) algorithm \cite{dmrg,dmrgbook}. In this method, successive
iterations are obtained by building larger (real space) blocks from smaller 
components. The Hamiltonian of each block can be diagonalized since in each 
iteration we truncate the states determined in the previous step. In the
truncation process, states in the small blocks are selected as a function
of their projection on the ground state of the larger block, and not as a 
function of their energies. The projection is computed from the
reduced density matrix of the smaller blocks. 
The iteration of this procedure allows to 
calculate ground state properties in disordered $1D$ systems with an accuracy 
comparable to exact diagonalization, but for much larger systems \cite{peter}.

It is important to recall that the eigenenergies of the many-body states
of the disordered ring need to be obtained with large precision since the
phase sensitivity is given as the difference of ground-state energies.
We can typically achieve system sizes corresponding to 50 particles on
100 lattice sites by keeping up to 2000 states per block, and then the
largest matrices that we have to diagonalize have a dimension of the
order of 10 millions.


\section{Half filling}
\label{sec:halff}
We will center our numerical studies on the charge density
and the $\Phi$-dependence of the ground state energy. We start in this section
by treating the case of half filling (the number of electrons is 
$N\!=\!M/2$), leaving the case of an arbitrary filling for section 
\ref{sec:deponfil}.

\subsection{Charge reorganization in individual samples}

\subsubsection{Charge density}

\begin{figure}[tb]
\centerline{\epsfxsize=\figwidth\epsffile{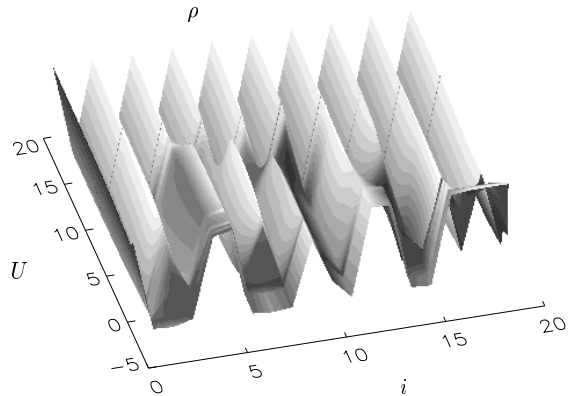}}
\vspace{2mm}
\caption[]{\label{density} Charge configuration for a typical sample 
(d of Fig.~\ref{gamma}) for $N\!=\!10$ particles on $M\!=\!20$ sites at 
$W\!=\!9$ as a function of the lattice site $i$ and the interaction
strength $U$.}
\end{figure}

 We start our analysis with
the reorganization of the ground state induced by the nearest-neighbor
(NN) repulsion (Fig.~\ref{density}), plotting the charge density $\rho$ 
(expectation value of $n_i$) as a function of $U$ and site 
index $i$ for a typical sample ($M\!=\!20$ and $N\!=\!10$). In order 
to favor the inhomogeneous configuration, the disorder is taken large 
($W\!=\!9$) such that the localization length (a rough guess 
is given by the perturbative result for weak disorder 
$\xi_1\!\approx\!100/W^2$) is of the order of the mean spacing 
$k_{\rm F}^{-1}\!=\!2$ between the charges. 
For $U\!\approx\!0$, one can see a strongly inhomogeneous density, while 
for large $U$ a periodic array of charges sets in. 
These two limits are separated by a sample-dependent crossover regime. 

\begin{figure}[tb]
\centerline{\epsfxsize=\figwidth\epsffile{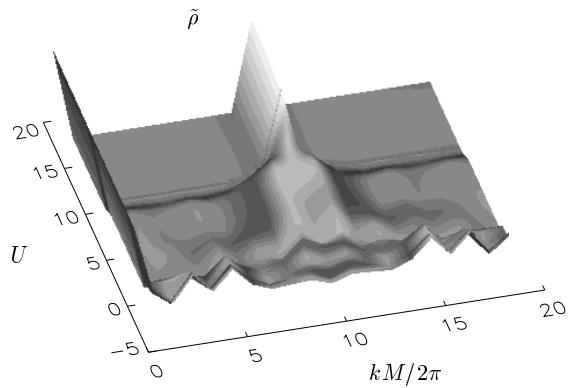}}
\vspace{2mm}
\caption[]{\label{fourier} Fourier transform of the charge configuration of
Fig.\ \ref{density}.}
\end{figure}

The charge reorganization can be clearly observed in the Fourier transform of
$\rho$ with respect to the space variable $i$ (Fig.~\ref{fourier}). For
weak interactions the Fourier transform does not show significant structure.
On the other hand, a single peak appears above a certain interaction
strength, reflecting the establishment of a regular periodic array of charges.
For intermediate interaction strengths, there is a tendency towards 
a periodic array, but defects persist and there are still several
Fourier components present.

\subsubsection{Density-Density correlation function}
\label{ddcf}

In order to quantitatively describe the sample-dependent reorganization 
of the charge density, we calculate \cite{wpsj} the density--density 
correlation function 
\begin{equation}
C(r)=\frac{1}{N}\sum_{i=1}^{M}\rho_i\rho_{i+r}
\end{equation}
for values $0\leq r \leq M/2$. The parameter 
$\gamma=\max_r\{C(r)\}-\min_r\{C(r)\}$
is used to distinguish between the electron liquid with constant density 
($\gamma\!=\!0$), and the regular crystalline array of charges 
($\gamma\!=\!1$). Since we include the translation $r\!=\!0$ in the 
definition of $\gamma$, we get $\gamma \neq 0$ for the electron
glass. Thus, $\gamma$ measures charge crystallization from an electron
liquid as well as the melting of the glassy state towards a more
liquid ground state.

\begin{figure}[tb]
\centerline{\epsfxsize=\figwidth\epsffile{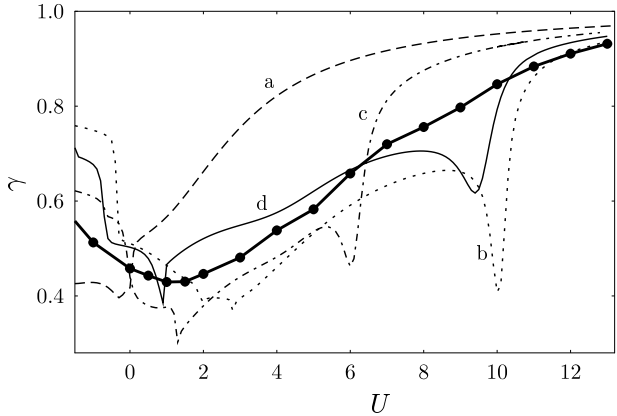}}
\vspace{3mm}
\caption[]{\label{gamma} Density--density correlation parameter $\gamma$ for 
four samples with $N\!=\!10$, $M\!=\!20$ and $W\!=\!9$. Thick dots:
Average over up to 166 samples.}
\end{figure}
 Fig.~\ref{gamma} shows the dependence of $\gamma$ on the interaction strength 
for four individual samples representing different behaviors. For certain 
impurity configurations, like in sample a, the periodic array is obtained at
a weak repulsive interaction, while one needs a strong interaction for other 
samples like b and d. Typically, $\gamma$ assumes a minimum for a small 
repulsive interaction at a sample-dependent value $U_{\rm c}$ of the
order of the kinetic energy scale $t$. This means
that the charge distribution is closest to a liquid around $U_{\rm c}$
and suggests a maximum of the mobility of the charge carriers. This is
an indication for a delocalization of the ground state by repulsive 
interactions.
In addition, most of the samples show small steps in the interval 
$0\le U \le 2t$, caused by instabilities between different configurations
of similar structure. The formation of the regular array of charges 
imposed by strong repulsive interactions occurs at a sample-dependent 
interaction strength $U_{\rm m}$. The step-like increase of $\gamma$ 
shows that the regular array is established abruptly at $U_{\rm m}$.

Typically the first and last dips of $\gamma$ for repulsive $U$,
signaling the above described charge reorganizations, are separated
by a transition region. In order to describe our problem as a 
transition between phases we have to consider the average behavior of
$\gamma$ and extract sample-independent values of the critical 
interaction strengths. The small dispersion of $U_{\rm c}$ yields an average 
$\gamma$ presenting a minimum at an interaction strength $U_{\rm F}$
of the order of $t$. 
This demonstrates the delocalization effect of a small repulsive
interaction accompanied by a ``more liquid'' charge density. The increase 
of the persistent current in $1D$ models with spin was traced back to the 
effect of repulsive interactions making the charge density more 
homogeneous \cite{GetS}. The present study shows that this mechanism 
also applies for spinless fermions in strongly disordered $1D$ chains.

Unlike the case of small $U$, for large interactions the jumps of
$\gamma$ are widely spread and therefore smeared out in the ensemble 
average. Thus, starting from the sample-dependent values $U_{\rm c}$ and
$U_{\rm m}$, it is possible to identify a sample-independent 
interaction strength $U_{\rm F}$ associated with the first charge 
reorganization by the minimum of $\langle \gamma \rangle$, but the 
last charge reorganization does not give 
a clear signature on $\langle \gamma \rangle$. In the next subsection we
give the estimation of the typical interaction strength needed to establish
the Mott phase. Recent works ~\cite{benenti_new,epl1,epl2} on $2D$ 
disordered clusters with Coulomb interaction also shows that one 
goes from the Fermi glass towards the pinned Wigner crystal through 
an intermediate regime (located between two 
different Coulomb-to-Fermi energy-ratios $r_s^{\rm f}$ and
$r_s^{\rm w}$). 
The difference with our problem is that, in the $2D$ case, 
a topological change can be observed in the pattern of the driven
currents, which cannot exist in the $1D$ case. However, in the two 
problems, we are addressing the difficult question of the quantum
melting of a solid array of charges (Mott insulator in our case,
Wigner crystal for $2D$ Coulomb repulsion) in the presence of a random
substrate. This melting can occur through a crossover regime 
(or an intermediate quantum phase in $2D$) where a liquid of
``defectons'' may co-exist with an underlying 
solid background, as suggested by Andreev and Lifschitz \cite{al}.

\subsubsection{Size dependence of the correlation parameter}
\label{defects}

The typical interaction strength needed to establish a perfectly regular
array of charges can be estimated from the competition between interaction
and potential energies in some special configurations. Starting from the
perfect Mott configuration with alternating charges, we see that we can 
gain potential energy by going to configurations
which are perfect only on parts of the lattice. The most favorable of 
this kind should typically be the case where the odd sites are occupied on 
half of the ring and the even sites on the other half of the ring, with two
domain walls between these regions. The cost in interaction energy is $U$,
while the gain in potential energy is of the order of $(W/2)\sqrt{N/2}$.

\begin{figure}[tb]
\centerline{\epsfxsize=\figwidth\epsffile{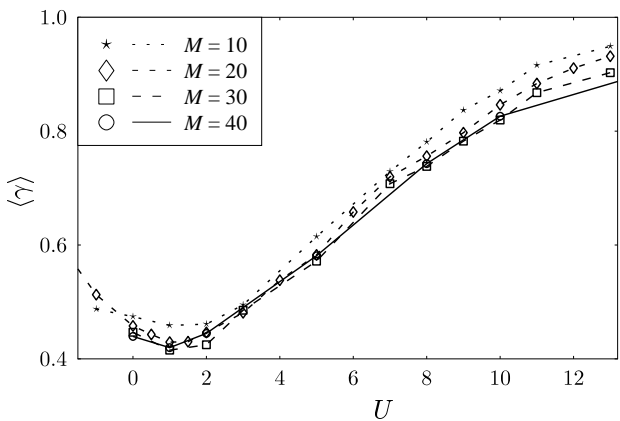}}
\vspace{3mm}
\caption[]{\label{gamma_size} Ensemble average of the density--density 
correlation parameter $\gamma$ as a function of the interaction strength for 
different system sizes $M$ at half filling and $W=9$.}
\end{figure}

The periodic array over the whole system is favored when the 
cost of the defect is larger than the gain in potential energy. This 
typically happens for 
\begin{equation}
U>U_{\rm W}=\frac{W}{2}\, \sqrt{N/2}\, .
\end{equation}
At half filling ($N=M/2$), the critical interaction strength $U_{\rm W}$
increases as the square root of the system size. 
In the dependence of the ensemble averaged correlation parameter on
$U$ for different system sizes (Fig.~\ref{gamma_size}) this tendency is
clearly visible. The interaction scale on which $\langle \gamma \rangle$ 
increases to its maximum value 1, characterizing the Mott insulator, 
is shifted to higher interaction values when the system size is increased. 
In the thermodynamic limit, one expects that no finite interaction strength 
(provided it is short ranged) can be sufficient to impose a perfectly 
ordered Mott insulator in a disordered system, consistent with the findings 
of Refs.~\cite{Shankar,imryma}.

\subsection{Phase sensitivity and localization}

In this section, we present calculations of the phase sensitivity of the
ground state which is defined, in Eq.~(\ref{eq:defphasen}), as the energy
difference between periodic ($\Phi\!=\!0$) and anti-periodic
($\Phi\!=\!\pi$) boundary conditions. 
This phase sensitivity is a measure of the localization of the electronic
wave-functions. The more insulating the system is,
the weaker will be the effect of boundary conditions. As discussed in
the introduction, the phase sensitivity conveys similar information, 
in the localized regime, as other measures of the response of the 
ground state to a flux threading the ring: the Kohn curvature
\cite{kohn} (charge stiffness) $\propto E^{''}(\Phi\!=\!0)$ and the 
persistent current $J \propto -E^{'} (\Phi)$. 

\subsubsection{Phase sensitivity in individual samples}

\begin{figure}[tb]
\centerline{\epsfxsize=\figwidth\epsffile{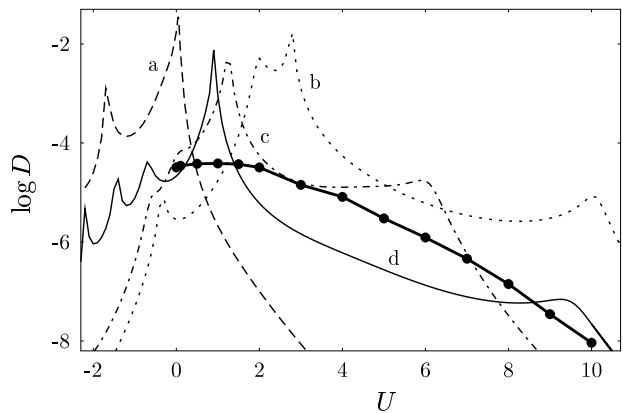}}
\vspace{3mm}
\caption[fig2]{\label{D(U)} Phase sensitivity $D(U)$ for four different 
samples with $N\!=\!10$, $M\!=\!20$, and $W\!=\!9$ in (decimal) logarithmic 
scale. Thick solid line and dots: average of $\log(D)$.}
\end{figure}

In Fig.~\ref{D(U)} we show the phase sensitivity $D(U)$ for the four 
samples presented in Fig.~\ref{gamma}, which were at half 
filling and with large disorder ($W\!=\!9$). Both for 
$U\!\approx\!0$ and $U\!\gg\!1$, $D(U)$ is very small, but sharp peaks appear 
at sample-dependent values $U_{\rm c}$, where the phase sensitivity can be  
{\em 4 orders of magnitude larger} than for free fermions. Remarkably, the 
curves for each sample do not present any singularity at $U\!=\!0$ which could 
have allowed to locate the free fermion case (that appears simply as an 
intermediate case). Peaks can be seen at positive and negative values of 
$U$.

It is important to notice that each peak of $D(U)$ in an individual sample
corresponds to a charge reorganization. This can be seen, directly, by 
following the evolution of the charge density as a function of $U$, or more 
systematically, by considering the one-to-one correspondence between the peaks 
of $D(U)$ and the dips of the density-density correlation parameter $\gamma$ 
(Fig.~\ref{gamma}). The information of $D(U)$ and $\gamma(U)$ is complementary,
but not equivalent: strong peaks of $D$ (happening for small values of $U$) 
correspond to small dips in $\gamma$, while the last charge rearrangement 
(leading to the Mott phase) is associated with a large dip in $\gamma$ and a 
small peak (or a shoulder) of $D$.

The free fermion case $U\!=\!0$ corresponds to an Anderson insulator. A small
repulsive interaction typically tends to delocalize the system and increases 
$D(U)$ until a charge reorganization takes place. In some cases (like 
sample a),
the first charge reorganization for positive $U$ immediately drives the system
to the homogeneous array of charges. In some other cases we go from the 
inhomogeneous density to the periodic array in a few steps signaled by 
additional peaks of the phase sensitivity. Examining the $U$ dependence of the 
density of those samples, one can note in some cases local defects in the 
periodic array subsisting up to large values of $U$. As discussed in 
Sec.~\ref{defects}, for a given interaction strength $U$, the appearance of 
defects becomes more and more likely as we increase the system size $M$ or the 
disorder $W$. Once the regular array of charges is established, the system 
becomes more and more rigid (pinned by the random lattice), and $D(U)$ 
decreases as a function of $U$. In Sec.~\ref{perturbation} we calculate, by 
perturbation theory, the phase sensitivity of the Mott insulator in a 
disordered potential.


\begin{figure}[tb]
\centerline{\epsfxsize=\figwidth\epsffile{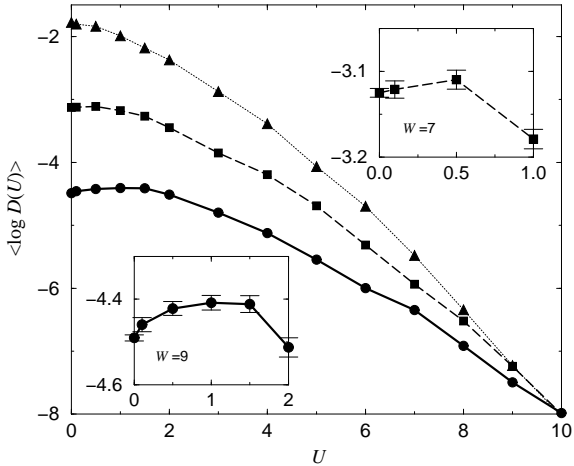}}
\vspace{3mm}
\caption[]{\label{average} Mean values of $\log D(U)$ for $M\!=\!20$, 
$N\!=\!10$ and three values of the disorder strength: $W\!=\!5$ (triangles), 
$W\!=\!7$ (squares) and $W\!=\!9$ (circles). Upper and lower insets: blow
up of the small-$U$ region for the last two cases showing a delocalization
effect.}
\end{figure}

\subsubsection{Mean values and statistics of the phase sensitivity}

\begin{figure}[tb]
\centerline{\epsfxsize=\figwidth\epsffile{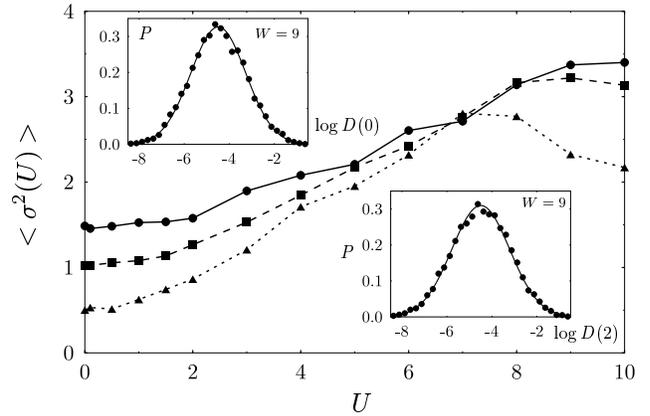}}
\vspace{3mm}
\caption[]{\label{variance} Variance of the phase sensitivity as a function of 
$U$ for three values of the disorder strength: $W\!=\!5$ (triangles),
$W\!=\!7$ (squares) and $W\!=\!9$ (circles). Upper and lower insets: 
probability distribution (dots) of $\log D(0)$ and $\log D(2)$ respectively, 
calculated from 10000 samples ($M\!=\!20, N\!=\!10, W\!=\!9$). The mean values 
and variances are $<\!\log D(0)\! >\;\approx -4.486$, 
$\sigma^2(0)\approx 1.49$ and $<\!\log D(2)\! >\;\approx -4.495$, 
$\sigma^2(2)\approx 1.66$.
The solid lines represent Gaussian distributions with these parameters.}
\end{figure}

The critical values $U_{\rm c}$ of the interaction strength are sample
dependent. 
Therefore, for a given value of $U$ we have a very different behavior for the
samples where $U$ is close to a critical value than for those where
it is not. Mixing the two situations leads to a widely fluctuating
distribution of $D(U)$ and if we average logarithms, as usually done 
in the localized regime, we obtain the smooth curve of Fig.~\ref{D(U)}.
We have checked that the probability distribution of D(U) is in fact log-normal
(see insets of Fig.~\ref{variance}).

In Fig.~\ref{average} we see that, for a given size and filling 
($M\!=\!20$, $N\!=\!10$), the log-average $\langle\log D(U)\rangle$ increases 
while decreasing the disorder ($W=9,7$ and $5$). For $U\!=\!0$ we have the
usual non-interacting behavior, while for large $U$ the phase sensitivity
becomes only weakly dependent on disorder (we will get back to this
point in Sec.~\ref{perturbation}). We also see that $\langle\log D(U)\rangle$ 
decreases with the interaction strength $U$, except for small $U$ 
($U\lesssim t$) 
and strong disorder ($W=9$ and $7$). The regions for $U\approx t$, for
$W=9$ and $7$, are blown up in the insets (lower left and upper right
respectively), and show that the interaction scale for the first charge
reorganization is weakly increasing with the disorder. The delocalization 
effect increases with the value of the disorder, but it is always very small. 
Our results for the average phase sensitivity are consistent with
those of Ref.~\cite{AetB}, finding a small increase of the average
persistent current in small $1D$ systems (up to 5 electrons on 10
sites) by direct diagonalization. We show in our work that such an
effect persists in larger systems and its small magnitude is to be 
contrasted with the spectacular enhancement found in individual 
samples.

The small delocalization effect on the average phase sensitivity
together with the large sample-to-sample fluctuations makes
it necessary to consider many impurity realizations (up to 5000) in
order to confirm the enhancement beyond the statistical uncertainty.
Comparing the information from Figs.~\ref{gamma} and 
\ref{average} we see that the average delocalization effect is more easily 
seen on the less fluctuating correlation parameter $\gamma$ than on $D$. 

The widely fluctuating values of $D(U)$ can be represented to a very
good approximation by a log-normal distribution for values of $U$ smaller 
than 8 (Fig.~\ref{variance}). The variances 
$\sigma^2(U)=\langle\delta(\log D(U))^2\rangle$ are of the order of 
$|\langle\log D(U)\rangle|$. They increase slowly with $U$ in the region of 
$U \lesssim t$ (roughly the interval over which we see the
delocalization effect), 
and more rapidly for larger values of $U$ (where the effect of the disorder 
becomes less important). Once the Mott regime is attained $\sigma^2$ no longer 
increases with $U$ (and becomes strongly dependent on $W$).

For a given sample the ratio between the phase sensitivity at a value of
$U\ne 0$ and at $U\!=\!0$ is also a widely fluctuating variable. Our numerical
simulations (not shown) demonstrate that the variable 
$\eta\!=\!\log (D(U)/D(0))$ is normally distributed \cite{sjwp}. 
Such a behavior does not simply follow from the log-normal 
distributions for $D(U)$ and $D(0)$, since these are not independent random 
variables. The width of the $\eta$-distribution is increasing with $U$, and 
for $U\!=\!2$ variations of $D$ over more than an order of magnitude are 
typical \cite{sjwp}. The samples a and b shown in 
Figs.~\ref{gamma} and ~\ref{D(U)} are characterized by extremely small
and large values of $\eta$, respectively, while c and d are
typical samples chosen around the center of the distribution.


\begin{figure}[tb]
\centerline{\epsfxsize=\figwidth\epsffile{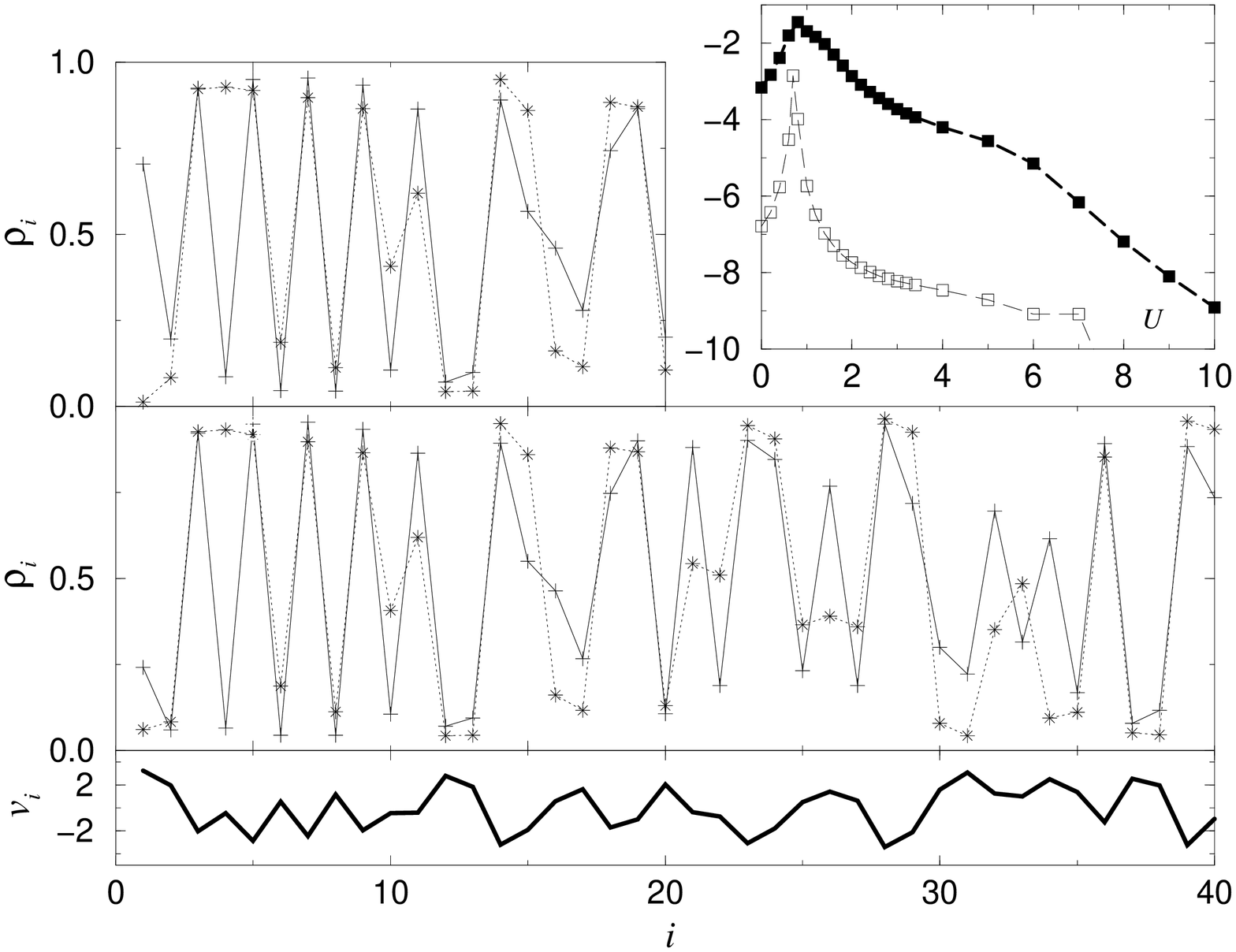}}
\vspace{3mm}
\caption[]{\label{doubling} Charge density as a function of the lattice site 
at half filling for a sample with $M\!=\!20$, $N\!=\!10$ (upper left panel) 
and with $M\!=\!40$, $N\!=\!20$ (central panel) with the impurity 
configuration $v_i$ represented by a thick solid line
in the lower panel ($W\!=\!7$). Stars and dotted lines correspond to 
$\rho_i(U\!=\!0)$, while pluses and solid lines to $\rho_i(U\!=\!3)$.
In the upper right panel the (decimal) logarithm of the charge sensitivity as 
a function of $U$ is shown for the two samples: $M\!=\!20$ (filled squares) 
and $M\!=\!40$ (empty squares).}
\end{figure}

\subsubsection{Size dependence and thermodynamic limit}

In the previous sections we described the increase of the phase sensitivity
as a signature of the charge reorganizations that eventually lead to a Mott 
insulator upon increasing the interaction strength. On the other hand, as 
discussed in Sec.~\ref{defects}, in the thermodynamic limit the disorder 
will lead to defects in the Mott phase at any
finite value of $U$. Two questions naturally arise at this point. Firstly, do 
charge reorganizations still exist and yield an enhancement of the phase 
sensitivity as we go to larger and larger systems? Secondly, will the 
delocalization effect survive in the thermodynamic limit?

In order to address the first question, we consider in Fig.~\ref{doubling} the
typical case of a sample with a strong disorder ($W=7$). The actual realization
of the impurity potential is sketched at the bottom panel (thick solid) for 
lattice sites going from $i\!=\!1$ to $i\!=\!40$. If we consider a sample of 
half such a size ($M\!=\!20$) at half filling ($N\!=\!10$) we obtain for the 
charge density (upper left panel) and for the phase sensitivity (filled 
squares, upper right panel) the kind of behavior previously discussed. A first 
charge reorganization takes place around $U=1$ where a peak in $D(U)$ is 
observed. If we now consider the whole sample ($M\!=\!40$) at the same 
filling ($N\!=\!20$), $\log D(0)$ as well as the overall values of $\log D(U)$ 
are reduced 
by a factor of $2$, but the charge reorganization involving the first half of 
the sample still takes place (medium panel) giving rise to a sharper peak of 
$D(U)$ at a slightly shifted critical value (empty squares, upper right
panel). This resonant-like behavior is generic. Upon increasing the
sample size the peaks of $D(U)$ become sharper (and slightly shifted), and
new charge reorganizations may yield supplementary structure in $D(U)$
(like the shoulder observed around $U\!=\!7$).

\begin{figure}[tb]
\centerline{\epsfxsize=\figwidth\epsffile{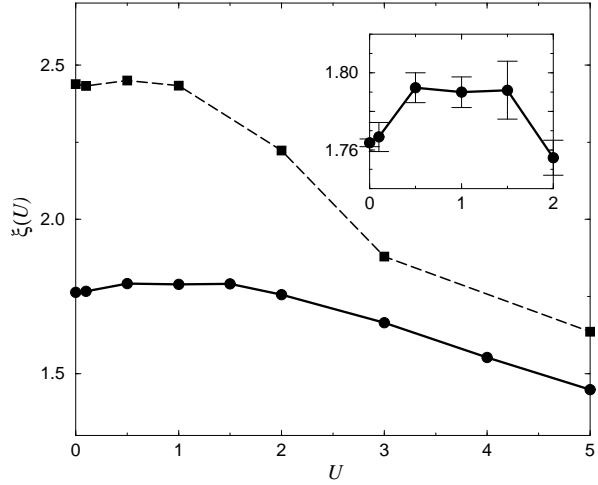}}
\vspace{3mm}
\caption[]{\label{loclen} Localization length obtained from the 
size-dependence of the phase sensitivity (Eq.~\ref{scaling}) as a function
of the interaction strength $U$ for half filling and two values of
the disorder: $W\!=\!7$ (squares) and $W\!=\!9$ (circles). Inset: blow up of 
the small-$U$ region for the case of $W\!=\!9$ showing a delocalization 
effect on $\xi$.}
\end{figure}

Having answered the first question for the affirmative we still need to 
settle the second one, since it is not obvious that the sharper peaks yield
a delocalization effect on the average. A detailed size-dependence study is 
needed, and we will focus on the interaction dependence of the localization
length. In the absence of interactions, a $1D$ disordered system is an
Anderson insulator, with the electron wave-functions localized
on the scale of the one-particle localization length $\xi_1$. 
Then, the persistent current and the phase sensitivity
scale with the system size $M$ proportional to 
$\exp{(-M/\xi_1)}$ \cite{Cheung}.

For our many-particle system, the (many-body) localization length $\xi (U)$
depends on the interaction strength and can be defined from the scaling
\begin{equation}\label{scaling}
\langle \ln D(U,M)\rangle = A(U) - \frac{M}{\xi (U)} \, ,
\end{equation}
which in the non-interacting case ($U \!=\!0$) yields the
standard one-particle localization length characterizing the spatial
decay of the electron wave-functions. The previous scaling is well
satisfied in the accessible space of parameters \cite{PhysicaE} and
allows to extract the values of $A(U)$ and $\xi (U)$. The former is
weakly dependent on $U$ for $0 \! \le \! U \! \le \! 5$ and more
strongly dependent for $U \! > \! 5$. The localization length
exhibits a similar behavior as $\langle \log D(U)\rangle$. It decreases 
with the interaction strength except for small $U$ and strong disorder, 
where a small delocalization is observed (Fig.~\ref{loclen}). For
$W\!=\!9$ (where the single-particle localization length is smaller 
than 2 lattice sites) a small repulsive interaction results in a $2\%$ 
effect on $\xi$, for $W\!=\!7$ the delocalization effect is smaller,
consistent with the behavior of $D(U)$. We therefore conclude that 
the delocalization on the average is not a finite-size effect.

\section{Physical Mechanisms}
\label{sec:phymec}

We have seen in the previous chapter how upon increasing the
strength of the electron-electron interaction we go from the
Anderson to the Mott insulating regimes through non-trivial
transitions. In particular, our numerical simulations show that the
intermediate regime is signed by charge reorganizations associated to 
an enhanced phase sensitivity. In this chapter we will try to
understand the mechanism underlying such behavior and we will 
develop a perturbation theory yielding the phase sensitivity in
the Mott regime in the presence of disorder.

\subsection{Simplified model of two particles on three sites}

The relation between charge reorganization and enhanced persistent
current can be understood in a simple toy model of two particles 
(spinless fermions) on three sites. We consider the Hamiltonian of 
Eq.~(\ref{hamiltonian}) with $M\!=\!3$ and switch off the 
interaction between the extreme sites 1 and 3. Therefore, we write
\begin{equation}
H_{(3)}=-t \sum_{i=1}^{3} (c_i^{\dagger} c_{i-1} + c^{\dagger}_{i-1}c_i) 
+\sum_{i=1}^{3}v_i n_i + U \sum_{i=2}^{3} n_i n_{i-1}\, .
\end{equation}
We keep the twisted boundary conditions, $c_0=\exp({\rm i}\Phi) c_3$ that 
allow us to address the phase sensitivity. The
site-dependent interaction mimics the fact that in the large-$M$ case
the sites 1 and 3 are joint through the rest of the chain (that we do
not include in the description of the present model). 

For two spinless fermions on three sites the Hilbert space has dimension 3 and 
the set $\{ |1,1,0\rangle, |1,0,1\rangle, |0,1,1\rangle \}$ is a convenient 
basis specifying the occupation of the sites. We can readily diagonalize our 
$3\times 3$ matrices for any value of our parameters $t$, $U$ and $v_i$.
However, in order to simulate the localized regime we will only consider the 
case where at least one of the on-site potentials $v_i$ is much larger than 
the kinetic energy scale $t$. We then study the transition from $U\!=\!0$ 
(where the potentials $v_i$ determine the charge distribution) to values of
$U$ much larger than all the $v_i$ (where the ground state is mainly
directed along the vector $|1,0,1\rangle$ in order for the electrons to 
avoid each other).

\subsubsection{Three-site model with fixed impurity configuration}

In a fist step we further simplify our problem by restricting the disorder to 
$v_1\!=\!v_2\!=\!-\epsilon$ and $v_3\!=\!\epsilon \! \gg \! t$. This 
configuration clearly favors the state $|1,1,0\rangle$ at $U\!=\!0$. In the 
two extreme cases the ground state energies are given by
\begin{subequations}\label{allgsab}
\begin{equation}\label{gsa}
E(U\!=\!0) \simeq \left(-2 \left(\frac{\epsilon}{t}\right) 
- \left(\frac{t}{\epsilon}\right) 
+ \frac{1}{2} \left(\frac{t}{\epsilon}\right)^2 \cos{\Phi}\right)t \, ,
\end{equation}
\begin{equation}
\label{gsb}
E(U\!\gg\!\epsilon) \simeq \left(-2 \left(\frac{t}{U}\right) 
- 2 \left(\frac{\epsilon t}{U^2}\right) 
+ 2 \left(\frac{t}{U}\right)^2 \cos{\Phi}\right)t \ ,
\end{equation}
\end{subequations}
The term $-2\epsilon$ in Eq.~(\ref{gsa}) is the on-site energy of the state 
$|1,1,0\rangle$, the next order terms in $t/\epsilon$ take into account the 
energy gain due to hopping. In the large-$U$ case there is no term in 
$\epsilon$ since the ground state is close to $|1,0,1\rangle$ and we have 
chosen $v_1\!=\!-v_3$. From Eqs.~(\ref{allgsab}) we can calculate the phase 
sensitivity in the two limiting cases and obtain
\begin{equation}\label{pstsm}
D(0) \simeq \frac{3}{2} \left(\frac{t}{\epsilon}\right)^2 t 
\gg D(U\!\gg\!\epsilon) \simeq 6 \left(\frac{t}{U}\right)^2 t \, ,
\end{equation}
in agreement with the fact that the phase sensitivity in the Mott phase is
reduced with respect to that of the Anderson phase. However, the decrease of 
$D$ with $U$ is not monotonous; for the critical value 
$U_{\rm c}\!=\!2\epsilon$ we 
have a degeneracy that leads to an increased phase sensitivity:
\begin{equation}
D(U_{\rm c}) \simeq \frac{3}{2} \left(\frac{t}{\epsilon}\right) t \gg D(0) \, .
\label{pstsmcu}
\end{equation}
The enhancement factor $\eta=D(U_{\rm c})/D(0)$ is then given by
\begin{equation}
\eta=\epsilon/t =U_{\rm c}/2t\, ,
\end{equation}
and even though both, $D(U_{\rm c})$ and $D(0)$ decrease when the disorder 
strength $\epsilon/t$ is increased, their ratio is increasing.  

Our toy model then captures the physics of a charge reorganization associated
with an enhanced sensitivity with respect to a perturbation at a point of 
degeneracy. The effect is reminiscent of the Coulomb blockade phenomenon. When 
the occupation numbers are approximately good quantum numbers (0 or 1) 
transport is blocked; in the transition between such extreme configurations the
occupation numbers are no longer good quantum numbers and transport is
favored. In the Coulomb blockade problem we have a true transport situation
and the degeneracy is between the state having $N$ electrons in the dot and
$1$ in the leads with the state having $N+1$ electrons in the dot. Moreover,
the constant charging energy model \cite{Been} allows us to think in terms
of a one-particle problem. In our case we do not have a transport configuration
but a delocalization of wave-functions and the degeneracy is between many-body
ground states of the ring.

\subsubsection{Disorder average in the three-site model}

The result above for a given potential realization can be used to determine 
the ensemble average of the critical interaction strength $U_{\rm c}$ within
this toy model. 

First of all, we relax the restriction to the disorder realizations,
allowing for arbitrary on-site energies $v_i \in [-W/2,W/2]$, assuming
box-distributions with probability density 
\begin{equation}
P(v)=\frac{1}{W} \,\Theta(W/2-|v|).
\end{equation}
For symmetry reasons, the exchange of the values $v_1\leftrightarrow v_3$ 
leaves $D(U)$ unchanged and we consider always $v_1>v_3$.

No level crossings between the ground state at $U=0$ (adapted to the 
disorder configuration), and the ground state at $U\gg W$ (close to
$|1,0,1\rangle$), occurs when the disorder realization favors the latter
already at $U=0$. This is the case if
\begin{equation}
v_2>v_1,v_3 \, ,
\end{equation} 
which is fulfilled in 1/3 of the whole parameter space.
Then, no positive $U_{\rm c}$ exists within our toy-model, and no
enhancement of $D(U)$ with respect to $D(0)$ can be expected for $U>0$.  

We now concentrate on the averaged enhancement factor 
$\langle \eta \rangle$, where the average is taken
over the parameter space $v_2<v_1$, in which an avoided level crossing occurs.
In the absence of hopping ($t=0$), the energies of the states $|0,1,1\rangle$ 
and $|1,0,1\rangle$ are given by $v_2+v_3+U$ and $v_1+v_3$, respectively, such
that a level crossing occurs at $U_{\rm c}=v_1-v_2$. Following the argument of
the previous section, the hopping then leads to a $D(U)$ which is enhanced at 
$U_{\rm c}$ by the factor $\eta=(v_1-v_2)/2t$. The average over the part of 
the parameter space $(v_1,v_2,v_3)$ in which $v_2,v_3<v_1$, using the box 
distributions yields 
\begin{equation}
\langle \eta \rangle =\frac{3}{8}\frac{W}{t}\, .
\end{equation}

The average enhancement of the phase sensitivity increases proportionally 
to the disorder strength, as the average peak position, 
$\langle U_{\rm c} \rangle =(3/4)W$. This behavior is consistent with
our numerical findings for larger systems.

\subsection{Avoided level crossings} 

In the toy model presented in the previous section we saw how an enhanced phase
sensitivity is linked with a charge reorganization of the ground state. The 
charge reorganization took there a very simple form: at a certain critical 
value of the interaction strength an electron jumps from one site to its 
neighbor. Or, more precisely, the ground state that was, for small $U$, mainly 
given by the vector $|1,1,0\rangle$ became, after $U_{\rm c}$, approximately aligned 
in the direction of $|1,0,1\rangle$. That can be seen as a crossing of two 
levels as a function of the parameter $U$. In this section we go back to our 
numerical simulations for large $M$ and demonstrate that the physics of level 
crossings is still valid despite the fact that the charge re-accommodation is 
not necessarily local ({\em i.e.} the electron may jump many sites across).
 
\begin{figure}[tb]
\centerline{\epsfxsize=\figwidth\epsffile{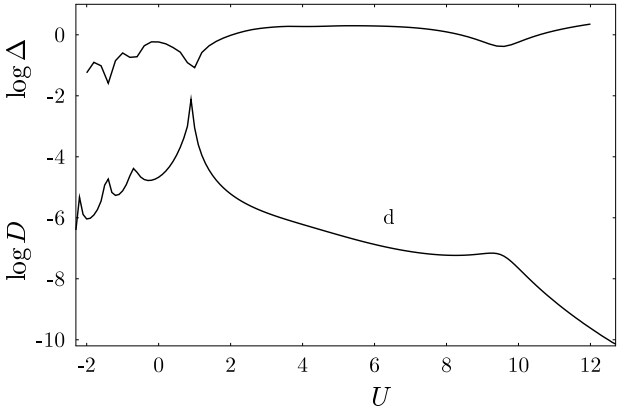}} 
\vspace{3mm}
\caption[]{\label{avcross} 
The energy spacing $\Delta$ between the ground state and the first 
excited state of sample d of Figs.~\ref{density},\ref{gamma} and \ref{D(U)} 
(upper line) together with the corresponding 
phase sensitivity (lower line).}
\end{figure}
The sharp changes in the ground state structure and the peaks observed in the 
phase sensitivity are the consequences of avoided crossings between the ground 
state and the first excitation, obtained upon increasing $U$.
While the ground state at weak interaction is well adapted to the disordered
potential, another state with a different structure and better adapted to 
repulsive interactions, becomes the ground state at stronger interaction. 
In the case of large disorder, with a one-particle localization length $\xi_1$ 
which is of the order of the mean distance between the particles, the overlap 
matrix elements between the different noninteracting eigenstates due 
to the interaction are very small and the levels almost cross. There is only 
a very small interaction range where a significant mixing (hybridization)
of two states is present. This is exactly where the peaks of the phase
sensitivity appear.

This scenario is confirmed by Fig.~\ref{avcross}, where the phase 
sensitivity $D$ and the energy level spacing $\Delta$ between the ground state 
and the first excited state are shown as a function of $U$ for sample d. 
Here, we used the Lanczos algorithm for direct diagonalization in order to 
obtain the energies of a few excited states.

Minima of $\Delta$ appear at the interaction values of the peaks  
in the phase sensitivity. A gap of increasing size opens between the ground 
state and the first excitation after the last avoided crossing at interactions 
$U>U_{\rm m}$, when the Mott insulator is established. A study of many other 
samples leads to the same conclusions. The statistics of the first excitation 
energy is therefore determining the behavior of the phase sensitivity.
In $2D$ with Coulomb repulsion, this has recently been addressed giving rise to
intermediate statistics~\cite{benenti_gap} at the opening of the quantum 
Coulomb gap. 

The importance of level hybridization in determining the persistent
current of many-particle systems has also been demonstrated in 
Ref.~\cite{CandB} within a slightly different model: A ring enclosing
a magnetic flux coupled to a side stub via a capacitive tunnel
junction. In particular, it was shown that the passage through the
hybridization point is associated with the displacement of charge in
real space, and also that strong enough Coulomb interactions can
isolate the ring from the stub, thereby increasing the persistent 
current.

      
\subsection{Phase sensitivity in the Mott insulator}\label{perturbation}

\subsubsection{Perturbation theory in $t/U$}
In the non-interacting limit, disorder leads to Anderson localization and 
the problem can be treated by an expansion in terms of $t/W$
\cite{bouch89}. In the
Mott insulator limit, the interaction dominates, and it is
possible to use an expansion in terms of $t/U$. In this second regime we 
decompose the Hamiltonian of Eq.~(\ref{hamiltonian}) as
\begin{equation}
H=H_0+H_1
\end{equation} 
with an unperturbed part containing disorder and interaction
\begin{equation}
H_0=\sum_{i=1}^{M} v_i n_i + U \sum_{i=1}^{M} n_i n_{i-1}
\end{equation}
and the perturbation given by the hopping term
\begin{equation}
H_1=-t \sum_{i=1}^{M} (c_i^{\dagger} c_{i-1} + c^{\dagger}_{i-1}c_i) \, .
\end{equation}
The solutions of the unperturbed part $H_0$ are simple. The 
eigenstates are products of on-site localized Wannier-states for each particle
\begin{equation}
|\psi_\alpha \rangle =\left(\prod_{k=1}^{N} c^\dagger_{i_k(\alpha)}\right) 
|0\rangle
\end{equation}
($|0\rangle $ is the vacuum state) and the corresponding eigenenergies are 
given by
\begin{equation}
E_\alpha = \sum_{k=1}^{N}v_{i_k(\alpha)} + U N_{\rm NN}(\alpha) \, . 
\end{equation}
$N_{\rm NN}(\alpha)$ is the number of particle-pairs in the
configuration $\alpha$ which occupy nearest-neighbor sites. 

It is important to notice that, in contrast to $H_0$, the perturbing part
$H_1$ depends on the boundary condition $c_0=\exp({\rm i}\Phi) c_M$.
For periodic (p, $\Phi=0$) and anti-periodic (ap, $\Phi=\pi$) boundary 
conditions one obtains 
\begin{equation}
H_1^{\rm p}=-t \sum_{i=2}^{M} (c_i^{\dagger} c_{i-1} + c^{\dagger}_{i-1}c_i) 
             - t (c_1^{\dagger} c_{M} + c^{\dagger}_{M}c_1) 
\end{equation}
and 
\begin{equation}
H_1^{\rm ap}=-t \sum_{i=2}^{M} (c_i^{\dagger} c_{i-1} + c^{\dagger}_{i-1}c_i) 
            + t (c_1^{\dagger} c_{M} + c^{\dagger}_{M}c_1) \, ,
\end{equation}
respectively.

The $n^{\rm th}$ order correction to the phase sensitivity is given by
\begin{equation}
D_n(U)=(-1)^N \ \frac{M}{2} \ (E^{\rm p}_n-E^{\rm ap}_n) \ ,
\end{equation}
where $E^{\rm p}_n$ and $E^{\rm ap}_n$ are the $n^{\rm th}$ order terms
in the perturbation expansion of the ground state energies for the 
Hamiltonians $H_1^{\rm p}$ and $H_1^{\rm ap}$, respectively.

\subsubsection{Phase sensitivity without disorder}  

The result for the problem without disorder has already been mentioned
in the literature \cite{tsiper}. 
We present here its derivation and generalize it to the case with disorder.
In the limit $t/U\!=\!0$, the ground state of our Hamiltonian 
at half filling is given by a regular array of charges without 
nearest-neighbor sites simultaneously occupied. There are two 
possibilities to realize such an array, either all the particles seat
on the odd sites of the chain 
\begin{equation}\label{gs_odd}
|\psi_0^{\rm o}\rangle =\left(\prod_{k=1}^{N} c^\dagger_{2k-1}\right) 
|0\rangle \ ,
\end{equation}
or the $N$ particles are on the even sites 
\begin{equation}\label{gs_even}
|\psi_0^{\rm e}\rangle =\left(\prod_{k=1}^{N} c^\dagger_{2k}\right) |0\rangle
\, .
\end{equation}
These states both have zero energy and therefore, in order to study
the effect of the boundary conditions, we have to use degenerate perturbation 
theory in the subspace spanned by $|\psi_0^{\rm o}\rangle$ and 
$|\psi_0^{\rm e}\rangle$. For each boundary condition, we can build
the perturbation expansion in $t$ from the matrix elements 
\begin{eqnarray}\label{pert_n}
&&{\cal H}_n^{s,s'}=\sum_{\alpha_1,\alpha_2,\dots ,\alpha_{n-1}}\times
\\ \nonumber
&&\frac{
\langle \psi_0^{\rm s} | H_1 | \psi_{\alpha_1} \rangle \langle \psi_{\alpha_1} 
|  H_1 | \psi_{\alpha_2} \rangle \dots 
\langle \psi_{\alpha_{n-1}} | H_1 | \psi_0^{\rm s'} \rangle }
{(E_0-E_{\alpha_1})(E_0-E_{\alpha_2})\dots (E_0-E_{\alpha_{n-1}})} \, ,
\end{eqnarray}  
with $s,s'=\{o,e\}$ and $H_1$ given by $H_1^{\rm p}$ or $H_1^{\rm ap}$
depending on the value of $\phi$ ($0$ or $\pi$, respectively). The sums 
run over all the eigenstates $\alpha$ of $H_0$ except the 
two degenerate basis states (\ref{gs_odd}) and (\ref{gs_even}).
There are two different types of matrix elements: 
the diagonal ones (${\cal H}_n^{e,e}$ and ${\cal H}_n^{o,o}$, starting and 
finishing at $|\psi_0^{\rm e}\rangle$ and $|\psi_0^{\rm o}\rangle$, 
respectively), and the off-diagonal ones (${\cal H}_n^{o,e}$ and 
${\cal H}_n^{e,o}$, starting and ending at different states). 

The numerators of Eq.~(\ref{pert_n}) contain matrix elements 
$\langle\psi_{\alpha_{i}} | H_1 |\psi_{\alpha_{i+1}}\rangle $ of the 
perturbing Hamiltonian. Since $H_1$ consists of one-particle hopping 
terms, non-zero matrix elements can arise only if the two states 
$|\psi_{\alpha_i}\rangle$ and $|\psi_{\alpha_{i+1}}\rangle$ differ by nothing 
else than the position of one of the particles. In addition, since the 
hopping terms allow only hopping to adjacent sites and we are dealing with 
spinless fermions, the order of the particles on the chain is conserved in 
the subsequent hoppings. 

\begin{figure}[tb]
\centerline{\epsfxsize=\figwidth\epsffile{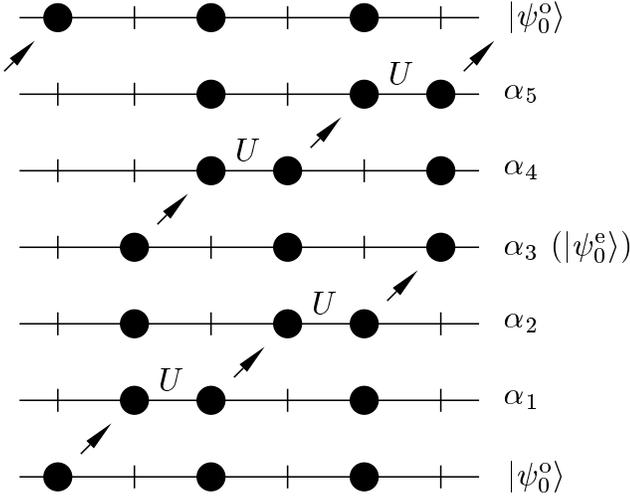}}
\vspace{3mm}
\caption[]{\label{sequence} A lowest order sequence contributing to
  the phase sensitivity for the example of $N=3$ and half filling. }
\end{figure}

In Fig.~\ref{sequence} we sketch a sequence of states $\alpha$ connecting
$|\psi_0^{\rm o}\rangle$ with itself for an $n\!=\!6$ diagonal contribution.
The first three steps of the sequence, connecting $|\psi_0^{\rm o}\rangle$
and $|\psi_0^{\rm e}\rangle$, represent an off-diagonal contribution to
$n\!=\!3$. We indicate with $U$ the interaction energy associated with each
of the intermediate states of the sequence. Generally, the sequences
in the states $\alpha$ can go ``forward'' and/or ``backwards''. For 
instance, a possible term contributing to
$n\!=\!2$ is that in which after the state $\alpha_1$ we go directly back
to $|\psi_0^{\rm o}\rangle$. However, since we are interested in the 
difference between periodic and anti-periodic boundary conditions, 
it is only the $\phi$-dependent terms that are relevant. 
That is, those involving a hopping between the sites $M$ and $1$.

In the case of periodic boundary conditions all the hopping terms have
a negative sign. Thus, the numerators appearing in Eq.~(\ref{pert_n})
are proportional to $(-t)^n$. For anti-periodic boundary conditions the
numerators are proportional to $(-t)^n(-1)^{h_{\rm b}}$, where $h_{\rm b}$ 
is the number of hoppings across the boundary between the sites $M$ and $1$.
Therefore, the corrections to the ground state energy due to the presence 
of finite hopping are the same for both boundary conditions, 
except for the contributions to the sums with odd $h_{\rm b}$. These last
contributions are the only ones relevant for the finite phase sensitivity.

The sequences with $n<N$ yield vanishing non-diagonal matrix elements,
since we need at least $N$ hoppings to go between $|\psi_0^{\rm o}\rangle$ 
and $|\psi_0^{\rm e}\rangle$. In the diagonal matrix elements with $n<M$ the 
sequences are such that each particle returns to its starting point, by doing 
forward and backward hoppings, and therefore $h_{\rm b}$ is necessarily
even. Thus, these diagonal matrix elements are independent on the 
boundary condition, and in addition ${\cal H}_n^{o,o}={\cal H}_n^{e,e}$
(the denominators do not depend on the boundary condition, nor on the
initial state).

The above described behavior of diagonal and non-diagonal matrix elements
shows that the degeneracy is not lifted for $n<N$. The lowest order in the
perturbation expansion which lifts the degeneracy, and at the same time
yields a contribution to the phase sensitivity, is $n=N$. For this order
of the perturbation we have finite non-diagonal matrix elements given
by sequences where each of the particles is moved by one site, the final
state being the other basis state of the degenerate subspace. The connection
between the two states $|\psi_0^{\rm o}\rangle$ and $|\psi_0^{\rm e}\rangle$ 
can be done by a sequence where all the particles hop forward, and also by a 
sequence of backward hoppings. If one of these two sequences crosses the 
boundary (yielding $h_{\rm b}=1$, changing the sign in the case of 
anti-periodic boundary conditions), the other does not 
(yielding $h_{\rm b}=0$).

For the sequences that go between $|\psi_0^{\rm o}\rangle$ and 
$|\psi_0^{\rm e}\rangle$ crossing the boundary, {\em either with periodic or
anti-periodic boundary conditions}, we have to consider an additional
sign $(-1)^{N-1}$ arising from the permutations needed to recover the
initial ordering of the fermionic operators in the final state. 
For an odd number of particles $N$ and anti-periodic
boundary conditions, the contributions of the forward and backward
sequences cancel each other and ${\cal H}_N^{o,e}={\cal H}_N^{e,o}=0$,
while with periodic boundary conditions both sequences add and we have
non-zero off-diagonal matrix elements that lift the degeneracy. For an 
even number of particles, the opposite behavior occurs:
The non-zero off-diagonal matrix elements are those corresponding to 
anti-periodic boundary conditions. In this way there is a difference between
periodic and anti-periodic boundary conditions for all possible values
of $N$. The corrections to the ground state energies are given by the 
lowest eigenvalue of the matrices ${\cal H}_N$. Since the diagonal
matrix elements are the same, and independent of the boundary conditions, 
for odd $N$ we have $E^{\rm p}_N-E^{\rm ap}_N < 0$, and for even $N$ we
have $E^{\rm p}_N-E^{\rm ap}_N > 0$. These signs ensure that $D_N$ is
positive, in agreement with a general theorem proposed by 
Leggett \cite{Leggett} and the result for a Luttinger liquid
\cite{loss}, fixing the sign of $E^{\rm p}-E^{\rm ap}$ according to
the parity of the number of electrons.

The splitting of the energy levels (and therewith the phase sensitivity) 
is given by the size of the off-diagonal matrix elements, which we estimate 
in the sequel. At half filling, the ground state configurations are the 
only ones which do not contain any nearest-neighbor sites simultaneously 
occupied, and their interaction energy vanishes. The eigenenergies of
the excited states of $H_0$ are $E_\alpha=U,2U,3U\dots$,
depending on the number of particles placed next to each other. Thus,
the absolute values of the denominators in Eq.~(\ref{pert_n}) are of the
order $U^{n-1}$. We have shown that the numerators are of order $t^n$, 
therefore, the parametric dependence of the lowest order ($N^{\rm th}$) 
correction to the phase sensitivity is given by
\begin{equation}\label{clean_lowest}
D_N(U) \propto U \left(\frac{t}{U}\right)^N
\end{equation}
with an $M$-dependent prefactor,
in agreement with the result mentioned by Tsiper and Efros \cite{tsiper}. 
Since higher order ($n>N$) terms contain higher powers of $t/U$, the result 
(\ref{clean_lowest}) is the leading correction in the limit of strong 
interaction $t/U\ll 1$.

\subsubsection{Phase sensitivity with disorder}     

The Mott insulator survives the introduction 
of disorder only for finite-size samples and not too strong disorder such that 
$U>W\sqrt{N}$. We will place ourselves in this regime in order to calculate 
perturbatively the phase sensitivity. The analysis presented for the clean 
case must be modified, and leads to a qualitatively different result. First 
of all, the two different possibilities to 
realize the regular array of charges will still be the energetically lowest 
configurations in the limit of strong interaction $U\gg t,W$, but the 
two states $|\psi_0^{\rm o}\rangle$ and $|\psi_0^{\rm e}\rangle$ are no 
longer degenerate. Their energies at $t\!=\!0$ are given by  
\begin{equation}
E_{\rm odd} = \sum_{k=1}^{N}v_{2k-1} 
\quad\quad \mbox{and} \quad\quad 
E_{\rm even} = \sum_{k=1}^{N}v_{2k}\, , 
\end{equation}
respectively, and differ typically by $W\sqrt{N}$. If this difference is much 
larger than their coupling due to the hopping terms (which, according to a 
perturbative argument like the one presented above, is of the order
of $t^N/U^{N-1}$), the ground state is given by one of the two base states,
but not by a superposition of them. For $N$ sufficiently large, or in
the limit of large $U$ where we are working, this is always true. 

\begin{figure}[tb]
\centerline{\epsfxsize=\figwidth\epsffile{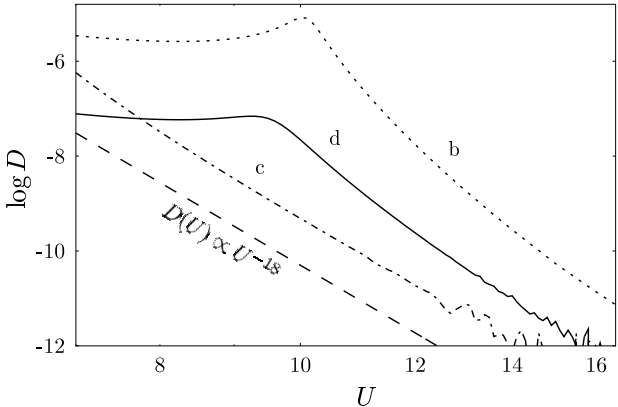}}
\vspace{3mm}
\caption[]{\label{large_u} The dependence of the phase sensitivity on 
the interaction strength for the samples b,c, and d of Figs.~\ref{gamma}
and \ref{D(U)}, at half filling ($N\!=\!10$, $M\!=\!20$, $W\!=\!9$), 
in double-logarithmic representation. 
Sample a exhibits the same asymptotic power law, but with a much
smaller prefactor such that the corresponding curve lies outside the
scale of the figure. The dashed line shows the perturbatively predicted slope 
$D(U)\propto 1/U^{2N-2}$. }
\end{figure}
As a consequence, one can use standard non-degenerate perturbation
theory and only sequences of hopping terms with $h_{\rm b}\ne 0$ which
connect the ground state to itself can give rise to a phase sensitivity. 
The lowest order contribution to the phase sensitivity must 
include now $M=2N$ hopping terms to translate each of the particles starting at
odd (or even) sites by two sites such that the final configuration is equal to
the initial one. This sequence contributes in Eq.~(\ref{pert_n}) to
the diagonal matrix element, connecting the ground state 
($|\psi_0^{\rm o}\rangle$ or $|\psi_0^{\rm e}\rangle$) with itself, 
proportionally to $(-t)^M (-1)^{N-1}$ in the case of periodic boundary
conditions. In the case of anti-periodic boundary conditions we must
add a sign $(-1)$ associated with the {\em one} traversal of the
boundary. In contrast to the clean case, forward and backward
sequences always cross the boundary and yield contributions of the
same sign to the diagonal matrix elements. Since the denominators do
not depend on the boundary conditions, the difference 
$E_M^{\rm p}-E_M^{\rm ap}$ has the correct sign \cite{Leggett} to
yield a positive $D_M$. 
  
As illustrated in Fig.~\ref{sequence}, one of the 
intermediate states can be the regular array with
the opposite parity than the initial state. The energy
difference associated to this state is not $U$ 
(like for all the other intermediate states with 
nearest-neighbor sites occupied) but $|E_{\rm odd}-E_{\rm even}|$.
The denominator containing the lowest 
power of $U$ is then of the order $W U^{2N-2}$ yielding
\begin{equation}\label{dirty_lowest}
D_{2N}(U) \propto U \left(\frac{U}{W}\right) \left(\frac{t}{U}\right)^{2N}
\, ,
\end{equation}
as the dominating behavior when $U \!\gg \!t,W$. This is
verified in our numerical simulations (Fig.~\ref{large_u}) when
considering individual samples after their last charge reorganization.
The weak dependence with respect to $W$ can be seen from the large-$U$
behavior in Fig.~\ref{average}.

\subsubsection{Localization length for the Mott insulator}
From the exponential size-dependence of Eqs.~(\ref{clean_lowest}) and
(\ref{dirty_lowest}), one can extract
the localization length in the Mott insulator. Imposing the size scaling
of Eq.~(\ref{scaling}) we obtain
for large $M$ a localization length
\begin{equation}
\xi_{\rm clean}=2\left(\ln \left(\frac{U}{t}\right)\right)^{-1}\, ,
\end{equation}
in the clean case, while with disorder we have
\begin{equation}
\xi_{\rm dirty}=1\left(\ln \left(\frac{U}{t}\right)\right)^{-1}\, .
\label{xidirty}
\end{equation}
Both typical lengths shrink with increasing interaction strength.
Interestingly, the presence of disorder reduces the localization length of the
Mott insulator to one half of the clean value, independent of the disorder
strength $W$. It is important to recall that the size-dependence that 
yields the typical length $\xi_{\rm dirty}$ of Eq.~(\ref{xidirty}) has to
be restricted to the condition $U>W\sqrt{N}>t^N/U^{N-1}$, allowing $N$ to 
vary over several orders of magnitude when $U$ is large. $\xi_{\rm dirty}$
describes the exponential decrease of $D$ with the system size within this 
range. 

The logarithmic dependence of the localization length on the
interaction parameter is also found for the Hubbard model in a clean
one-dimensional system \cite{Staff93} as well as in a disordered $2D$
model with Coulomb repulsion \cite{SetW}.

      
\section{Dependence on filling}
\label{sec:deponfil}

In the previous sections we have been mainly concerned with the
interaction effects of spinless fermions at half filling. At 
half filling, and with strong disorder, a short-range interaction has
a dramatic effect on the charge density. We have seen that
reducing the disorder increases the localization length and
results in less important charge reorganizations. Reducing the
electron density increases the inter-particle distance, making a
short-range interaction less effective. In this section we 
analyze the effect of short-range interactions as we move away
from the optimal conditions of strong disorder and half filling.

\begin{figure}[tb]
\centerline{\epsfxsize=\figwidth\epsffile{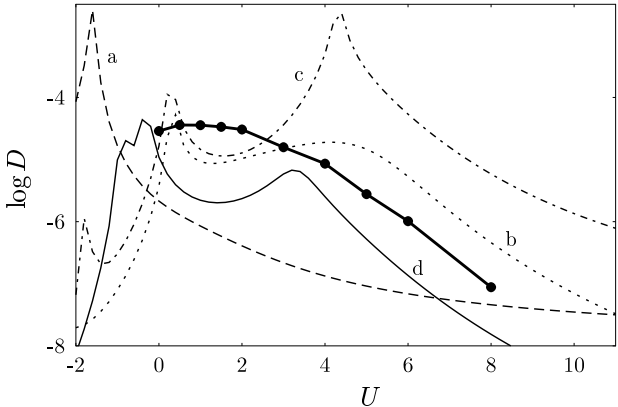}}
\vspace{3mm}
\caption{Phase sensitivity $D(U)$ for four samples with the
impurity configurations of those of Fig.~\ref{D(U)}, but with 
only $N\!=\!9$ particles ($M\!=\!20$, $W\!=\!9$). The average 
over many samples is represented by the thick dots.}
\label{N=9}
\end{figure}

\subsection{Fillings close to one half}

In Fig.~\ref{N=9} we present the phase sensitivity for strong
disorder ($W\!=\!9$) and a filling close to one half ($N\!=\!9$, 
$M\!=\!20$). In analogy with the case of half filling presented in
Fig.~\ref{D(U)}, we see strong peaks of $D$ for individual
samples and a broad maximum for the average values (thick dots). 
We therefore confirm that the physics of charge reorganizations 
is robust, and it is not a simple commensurability effect restricted to 
half filling.

\subsection{Phase sensitivity in the large-$U$ limit}
\label{subsec:pslUl}

The similarities between Figs.~\ref{D(U)} and \ref{N=9}, that we have
pointed out above, concern the charge reorganizations and the corresponding
peaks of $D$ for relatively weak interactions (of the order of
$t$). In the large-$U$ limit
there appear some differences between the cases of half filling and 
less than half filling. For instance, we can see a tendency towards
saturation of $D(U)$ for the samples a and c of Fig.~\ref{N=9}. In
fact, in a larger $U$-range we observe that all the curves for
densities less than one half saturate for sufficiently large $U$.

This saturation is easy to understand from the perturbative analysis 
given in Section~\ref{perturbation}.
In the sequence of intermediate states $|\psi_\alpha \rangle$
contributing to each of the terms in Eq.~(\ref{pert_n}) (one of them
represented in Fig.~\ref{sequence}) we readily see that at
less than half filling it becomes possible to move the particles one 
after the other on the chain without ever having two particles on 
neighboring sites. These sequences give boundary condition dependent
contributions to $E_n$ without any dependence on $U$ in the
denominators $(E_0-E_{\alpha_i})$. Since they are the only
contributions which survive in the large-$U$ limit, they
define the value at which the phase sensitivity saturates. 

The same happens at more than half filling, when already the ground state 
contains the minimum number $2N-M$ of nearest neighbors. Here, it is possible 
to translate all of the particles at constant interaction energy thereby 
leaving $U$-independent contributions to (\ref{pert_n}).   
 

\subsection{Charge reorganization at high disorder}

Repulsive interactions give rise to charge reorganizations when the
non-interacting ground state fixed by the disorder configuration is
not well adapted (energetically) to a finite value of $U$.
Assuming a very strong random potential, such that the 
one-particle states are close to on-site Wannier states, the ground state 
of $N$ particles on $M\ge 2N$ sites (more than half filling can be treated 
similarly, using the particle-hole symmetry) at $U=0$ is given by the 
particles occupying the $N$ lowest sites. The positions of these $N$ sites on 
the ring are random. No reorganization of the ground state due to short-range
repulsive interaction takes place when this configuration does not contain any 
two particles on neighboring sites. 

The probability for obtaining such a configuration can be calculated as 
follows. One starts from the configuration where the particles occupy the
$N$ first odd sites of the ring $\{1,3,\dots ,2N-1\}$ and each of the 
particles has `his' empty even site on its right. Then, $M-2N$ empty sites are 
still available which can be distributed among the $N$ gaps on the right hand 
side of the particles, and we obtain
\begin{equation}
\Omega=\frac{M}{N}\left(\begin{array}{c}M-N-1\\M-2N\end{array}\right)
\end{equation}
different possible configurations without nearest neighbors. The factor of
$M/N$ arises because of the $M$-fold translational symmetry of the problem and 
the indistinguishability of the $N$ particles. 

Since the total number of possibilities to place $N$ spinless fermions on $M$ 
sites is given by 
\begin{equation}
T=\left(\begin{array}{c}M\\N\end{array}\right)\, ,
\end{equation}
the probability to have a configuration without nearest neighbors in a given 
sample is 
\begin{equation}
\label{eq:probexac}
P=\frac{\Omega}{T}=\frac{M}{N}
\frac{\left(\begin{array}{c}M-N-1\\M-2N\end{array}\right)}{
\left(\begin{array}{c}M\\N\end{array}\right)}\, .
\end{equation}
Introducing the filling factor $x=N/M$, and using Stirling's approximation 
for the evaluation of the factorials, we find
\begin{equation}
\label{eq:probapprox}
P \simeq e^{M g(x)} 
\quad\quad \mbox{with} \quad\quad
g(x)=\ln{\left(\frac{(1-x)^{2(1-x)}}{(1-2x)^{(1-2x)}}\right)}\, .
\end{equation} 
Since $g(x)$ is negative for all $0<x\le 1/2$, we obtain always $P=0$
in the thermodynamic limit $M\rightarrow\infty$ at constant filling $x$.
Therefore, {\em in the case of strong disorder} ($W \gg t$), the probability 
to obtain a charge reorganization due to repulsive 
interactions is one in the thermodynamic limit, at arbitrary filling.

\begin{figure}[tb]
\centerline{\epsfxsize=\figwidth\epsffile{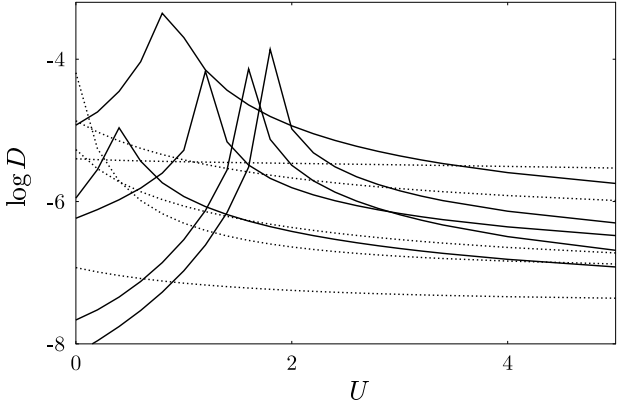}}
\vspace{3mm}
\caption{Phase sensitivity for ten samples with quarter-filling
($N\!=\!5$, $M\!=\!20$) and strong disorder ($W\!=\!9$). Roughly half
of the samples exhibit a peak in $D(U)$ (solid lines), while the other
half yield a monotonously decreasing phase sensitivity.}
\label{qfW=9}
\end{figure}
In Fig.~\ref{qfW=9} we consider the case of strong disorder ($W\!=\!9$) and 
quarter filling ($M\!=\!20$, $N\!=\!5$). For large $U$ we observe the
saturation of $D(U)$ described in the previous chapter. In the regime
of smaller $U>0$ we observe samples for which there is a peak in $D(U)$
(exhibiting also a charge reorganization) and others in which $D(U)$ is
a monotonously decreasing function. In the former case the $U\!=\!0$
configuration has nearest-neighbor sites that are occupied, while
in the latter there is no occupancy of nearest neighbors in the 
non-interacting problem, and by increasing $U$ we do not achieve 
any substantial charge reorganization. A systematic study over many
samples with the conditions of Fig.~\ref{qfW=9} yields a probability
of about 0.5 to obtain a sample which does not present a peak in $D(U)$,
while Eq.~(\ref{eq:probexac}) predicts $P \approx 0.26$. This discrepancy
can be traced to the fact that Eqs.~(\ref{eq:probexac}) and 
(\ref{eq:probapprox}) are valid in the limit of a very strong 
disorder $W \gg t$. For the parameters used in Fig.~\ref{qfW=9} it
is not always true that the $N$ particles occupy the $N$ sites with
lower electrostatic potential, since the non-zero kinetic energy can
split the two levels associated with two neighboring sites if their
on-site energies are closer than $t$ ({\em i.e.} when a local 
potential well is formed by two nearly-degenerate sites). This splitting
reduces the probability to have two adjacent sites occupied in the
non-interacting ground-state.
 
    
\subsection{Low fillings - weak disorder}

We have analyzed above the limit of very strong disorder, when the 
one-particle states are almost completely localized on one of the sites. 
We have seen that in this regime a short-range interaction leads 
to a charge reorganization whenever two neighboring sites are occupied 
at $U\!=\!0$. Therefore, in finite size samples, the probability of having 
peaks in $D(U)$ is 
maximum at half filling. At weaker disorder, and without interaction,
the particles are localized over several sites. The occupation of
a site is no longer an almost good quantum number (0 or 1) and turning
on a short-range interaction does not always result in a charge
reorganization.

\begin{figure}[tb]
\centerline{\epsfxsize=\figwidth\epsffile{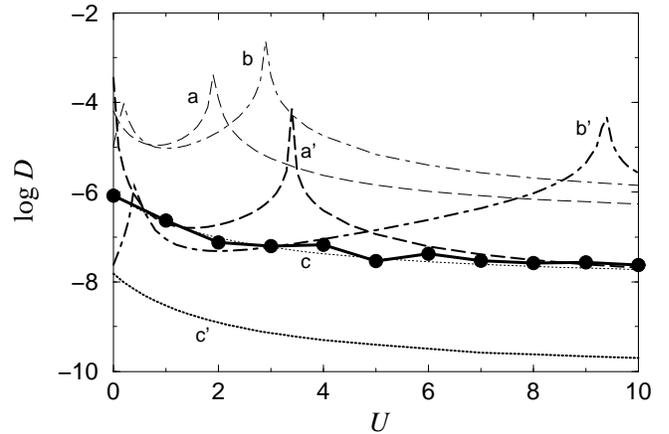}}
\vspace{3mm}
\caption{Thin lines: three samples (a, b and c) with $M\!=\!40$,
$N\!=\!10$ and $W\!=\!4$. Thick lines: three samples (a', b' and c') with
$M\!=\!40$, and $N\!=\!10$ obtained by scaling the previous impurity potentials
from $W\!=\!4$ to $W\!=\!5$. Thick solid line and dots: average of $\log(D)$
for $W\!=\!5$.}
\label{quarterf}
\end{figure}
In Fig.~\ref{quarterf} we present the case of quarter filling
($N\!=\!10$, $M\!=\!40$) for three samples (a, b and c) with a
disorder $W\!=\!4$, and another three samples (a', b' and c')
obtained from the first ones by scaling the impurity potentials to
$W\!=\!5$.  For instance, the configurations of samples a and a'
are obtained from the same set of random numbers, and it is only the
overall scale of the fluctuations that changes. For certain samples
(like c) the non-interacting system has all the electrons localized
far away from each other and the charge configuration does not change
when turning on the interactions, resulting in a monotonously
decreasing $D(U)$. On the other hand, in other samples (like a and b)
the non-interacting one-particle wave functions overlap considerably 
and the appearance of a short-range interaction results in a charge
reorganization. In all cases we appreciate the saturation of $D(U)$
for large $U$, as discussed in \ref{subsec:pslUl}.

Going from $W\!=\!4$ to $W\!=\!5$ for each of the samples changes
the details of the curves, but not their nature. For instance, the
peaks of a' and b' are obtained for stronger
interactions than those of a and b. Such a behavior is 
understandable since higher disorder within the same random
configuration necessitates a stronger interaction to produce a
charge reorganization. If we keep increasing $W$ for a sample
exhibiting a peak, we obtain sharper resonances at increasing
values of the interaction. Of course, for all samples, the typical 
values of $D$ are reduced when increasing $W$. Decreasing the
disorder below $W\!=\!4$ results in broader peaks, that disappear
once the localization domain of the individual wave-functions becomes of
the order of the inter-particle distance. The presence of peaks in
the phase sensitivity is therefore fixed by the impurity configuration
of the sample, and independent of the strength of the disorder in some 
range of $W$.

The different behaviors obtained among the samples with a given value $W$ of
the disorder are responsible for the fact that the average phase
sensitivity ($\langle \log D(U) \rangle$, thick dots in  Fig.~\ref{quarterf})
decreases monotonously with the interaction strength, instead of exhibiting
the broad maximum obtained for half filling and high disorder. We have 
seen that in the limit of very strong disorder it was possible to give a 
crude estimation of the probability to have a sample that presents a peak
in $D(U)$. Once the disorder is weak enough to have single-particle
states localized over several sites, we know that such a probability 
decreases, but it is rather difficult to extend the previous analysis in 
order to give an estimation of it. The relevant parameters are the
one-particle localization length $\xi_1$, fixed by the disorder, and the
inter-particle distance $1/k_{\rm F}$, fixed by the electron filling. As a
function of these two parameters we can clearly distinguish two cases: 

(i) When $\xi_1$ becomes much smaller than $1/k_{\rm F}$, the charge
density without interaction in finite size samples is more and 
more likely to consist of $N$ distant peaks. Since such configurations 
are well adapted to the interacting case as well, the probability to find 
reorganizations of the ground state due to the interaction is reduced.

(ii) In the opposite case, when $\xi_1$ becomes much larger than
$1/k_{\rm F}$, no distinct one-particle peaks are present in the charge
density at $U\!=\!0$, and there are no regions of the sample with very
small charge density. While increasing $U$, the total energy is minimized by
gradually pushing the particles away, but there are no sudden charge
reorganizations accompanied with peaks of the phase sensitivity.  
This behavior leads to a smooth decrease of $D(U)$.

Between the two previous cases, we find the optimal situation,
$k_{\rm F} \xi_1 \approx 1$, to observe the delocalization effect of interactions. 
This rough criterium is based on mean values (filling $N/M$ and disorder $W$) 
and only fixes the probability to obtain peaks in $D(U)$. The occurrence,
or not, of charge reorganizations depends on the specific sample. That is,
on the random configuration of the impurity potential. We
have seen that if a given sample exhibits a peak of $D(U)$, such a
behavior is maintained over some range of values of the disorder $W$
(small enough not to take us away from the condition 
$k_{\rm F} \xi_1 \approx 1$ and
into the limits (i) and (ii) previously discussed). It is worth to notice that
the disappearance of charge reorganizations, that we obtain upon
decreasing the disorder or the electron density, are tightened to the
fact that we work with repulsive nearest-neighbor interactions. A long
range interaction could be effective in yielding charge reorganizations
away from the optimal conditions of half filling and strong disorder. 


\section{Conclusions}
\label{sec:conclusion}

We have investigated spinless fermions in strongly disordered chains, 
as a function of the strength of a short-range repulsive interaction, 
mainly for densities corresponding to half filling. Such a system is
a Fermi glass (Anderson insulator) for small values of the interaction
(when the disorder is the dominant energy scale), and a Mott insulator
when the interaction dominates over the disorder. 
Using a powerful numerical method, the density matrix 
renormalization group algorithm, we have been able to address the
transition regime between these two previous limits, which is not
accessible by perturbation theory or mean field approaches. We have calculated 
the charge density of the many body ground state, as well as the dependence 
of the ground state energy on the boundary condition. This last property,
quantified by the so-called phase sensitivity, is related
to the transport properties and also to the persistent
current in the chain. It has a small finite value 
for the Anderson insulator and decreases with increasing interaction 
strength in the Mott insulator. 

In striking contrast to the case of weak disorder, where
repulsive interactions always strengthen localization for
spinless fermions in one dimension, we have shown that in strongly 
disordered samples, and close to half filling, the ensemble average of the
phase sensitivity can be enhanced by a repulsive interaction $U$.
It shows a maximum at a value $U_{\rm F}\approx t$, that is, for 
interaction strengths of the order of the kinetic energy scale. 
The study of the density-density correlations reveals that the ground
state is the most homogeneous (liquid-like) for $U \approx U_{\rm F}$, 
consistently with the maximum in the phase sensitivity. 

An analysis of different system sizes shows that these features
persist in the thermodynamic limit. Even though the system stays an
insulator for all values of the interaction strength, the size dependence 
of the phase sensitivity allows to extract a many-body localization
length, which is slightly enhanced by the interactions and maximum at 
$U_{\rm F}$. 

While the interaction-induced enhancement of the phase sensitivity is a 
rather small effect for the ensemble average, it can reach several orders 
of magnitude in individual samples, at sample-dependent values $U_{\rm c}$.
At these values of the interaction strength, abrupt reorganizations of the 
many-body ground state structure occur. The transition from the Anderson 
insulator to the Mott insulator, upon increasing the strength of the
repulsive interaction, is typically made in two steps. In a finite size 
system, the first change of the ground state structure at $U_{\rm c}$ is 
followed, at much larger interaction values $U_{\rm m}\propto W$, 
by the installation of the Mott insulator. The interaction strength 
$U_{\rm m}$ diverges in the thermodynamic limit, reflecting the fact that
the perfect order of the Mott insulator is destroyed by an arbitrarily 
small amount of disorder in an infinite size system.
We have found that the reorganizations of the 
ground state structure correspond to avoided level crossings between 
the many body ground state and the first excited state, as a function
of the interaction strength. We have used simple toy-models to clarify
the relationship between avoided crossings, charge reorganizations, and
peaks in the phase sensitivity.

The two limits of weak and strong interaction can be understood using 
perturbative schemes. Recently published Hartree-Fock results 
\cite{kambili} exhibit some qualitative similarities with our quasi-exact
results. However, the average phase sensitivity is strongly
underestimated by Hartree-Fock at strong interaction values. A
direct comparison of the phase sensitivity for individual samples
\cite{montambaux} shows that Hartree-Fock is
rather accurate for weak interaction values up to the first charge 
reorganization ($U<U_{\rm c}$), but fails for
stronger interactions when the
ground state is very different from the non-interacting ground state. 
In the regime of very strong interaction, a perturbative development 
starting from the Mott insulator
can describe quantitatively the phase sensitivity for $U>U_{\rm m}$. 

In contrast, the intermediate regime $U_{\rm F}<U<U_{\rm W}$ where
both, interaction and disorder play an important role, is intrinsically
non-perturbative and more difficult to analyze. The interplay of
interactions and disorder leads to strong electronic correlations in
the intermediate regime, resulting in a behavior qualitatively
different from the two limiting situations.

In the limit of strong disorder, we were able to provide a rough
estimate of the probability of observing a charge reorganization in a
given sample, as a function of the particle density and the system
size. This shows that for strong disorder, half filling is the optimal
condition to obtain interaction-induced charge reorganizations. 
At lower electronic densities, the probability in finite-size samples 
is reduced with respect to this optimal situation. 
However, when the disorder strength is reduced 
simultaneously, such that the one-particle localization length is of
the order of the inter-particle distance, the reduction of the probability
is less important. We expect that the use of a long-range interaction
could also favor the occurrence of charge reorganizations.

We recall the fact that the delocalization effect of repulsive
interactions has been predicted in models other than that of our work:
Disordered Hubbard models in $1D$ \cite{GetS} and in $2D$ \cite{KDS}, systems
with strong binary disorder \cite{ulmke}, rings coupled to a side stub
\cite{CandB}, and interacting bosons in a disordered chain \cite{Sca}.

As discussed in the introduction, the understanding of the metal-insulator
transition in disordered two-dimensional systems has been one of our
motivations for this study. Even if our model is much simpler than the
realistic problem of interest, it contains non-trivial features that may
be useful to understand the transition, like the concept of charge
reorganization discussed above. The intermediate regime 
that we find between the limits of weak and strong interactions can be 
considered as a precursor of the correlated phase found in numerical 
studies \cite{benenti_new} of two dimensional disordered clusters 
(with long-range Coulomb interaction) when the Wigner molecule 
is about to be formed. Consistently with our results, the use of the
Kubo-Greenwood formula and an exact diagonalization in a truncated
basis of Hartree-Fock states \cite{Schr} yields an average conductance
which is slightly increased by a small repulsive interaction for
spinless electrons in strongly disordered $2D$ systems.

Recent experimental measurements of the local compressibility in the 
localized phase of a two-dimensional system \cite{ilani} yielded 
important spatial inhomogeneities and very large fluctuations as a
function of the carrier density, while the metallic phase appeared as
spatially homogeneous and less fluctuating. The strong fluctuations of
the localized phase can be interpreted as charge reorganizations, very 
much in line with the interpretation of peaks in the phase sensitivity 
that we have thoroughly discussed in our work.

Since the existing numerical work in two dimensions relies on some drastic
approximations or is restricted to small systems, it could be useful
to extend our numerical techniques beyond the one-dimensional case. In
addition, the electron spin has been shown experimentally to play a major
role in the metal-insulator transition \cite{Pepper,expspin}, and it should be
interesting to relax the condition of spinless fermions in order to
approach the realistic case.

 
\subsection*{Acknowledgements}
We thank G.\ Benenti, G.-L.\ Ingold, A.\ Kampf, D.\ Vollhardt, 
and X.\ Waintal for useful discussions, and Ph.\ Jacquod for drawing
our attention to Ref.\ \cite{Shankar}.
We gratefully acknowledge financial support from the DAAD and the A.P.A.P.E. 
through the PROCOPE program and from the European Union through the TMR 
program. 


\begin{thebibliography}{99} 

\bibitem{A&M} N.W.~Ashcroft and N.D.~Mermin, 
              {\em Solid State Physics},
              (Saunders College, Philadelphia, 1976).

\bibitem{levy90} L.P.~L\'evy, G.~Dolan, J.~Dunsmuir, and H.~Bouchiat,
                 Phys.\ Rev.\ Lett.\ {\bf 64}, 2074 (1990).

\bibitem{Webb1} V.~Chandrasekhar, R.A.~Webb, M.J.~Brady, M.B.~Ketchen,
                W.J.~Gallagher, and A.~Kleinsasser,
                Phys.\ Rev.\ Lett.\ {\bf 67}, 3578 (1991).

\bibitem{Webb2} P.~Mohanty, E.M.Q.~Jariwala, M.B.~Ketchen, and R.A.~Webb, 
                in {\em Quantum Coherence and Decoherence},
                edited by K.~Fujikawa and Y.A.~Ono (Elsevier, 1996).

\bibitem{ensemble} A.~Schmid, 
                   Phys.\ Rev.\ Lett.\ {\bf 66}, 80 (1991);
                   F.~von Oppen and E.K.~Riedel, 
                   {\it ibid} 84; 
                   B.L.~Altshuler, Y.~Gefen, and Y.~Imry, 
                   {\it ibid} 88.

\bibitem{Eckern} U.~Eckern, 
                 Z.~Phys.\ {\bf B 42}, 389 (1991).

\bibitem{ESra} U.~Eckern and P.~Schwab, 
               Adv.~Phys.~{\bf 44}, 387 (1995).

\bibitem{Kravchenko} S.V.~Kravchenko et al., 
                    Phys.\ Rev.\ B {\bf 50}, 8039 (1994); 
                    {\it ibid} {\bf 51}, 7038 (1995).

\bibitem{Pepper} A.R.~Hamilton {\em et al.}, Phys.\ Rev.\ Lett.\ 
                {\bf 82}, 1542 (1999).

\bibitem{SiGe} J.~Lam {\em et al.}, Phys.\ Rev.\ B {\bf 56} R12741 (1997);
               P.T.\ Coleridge, {\it ibid}, R12764. 

\bibitem{Go4} E.~Abrahams, P.W.~Anderson, D.C.~Licciardello, and
                T.V.~Ramakrishman, Phys.\ Rev.\ Lett.\ {\bf 42}, 673 (1979).

\bibitem{abrahams} E.~Abrahams, S.V.~Kravchenko, and M.P.~Sarachik, 
                   cond-mat/0006055.

\bibitem{benenti_new} G.\ Benenti, X.\ Waintal, and J.-L.\ Pichard, 
                      Phys.\ Rev.\ Lett.\ {\bf 83}, 1826 (1999).

\bibitem{spivak} B.~Spivak, cond-mat/0005328

\bibitem{Moha} P.~Mohanty, E.M.Q.~Jariwala, and R.A.~Webb, 
                Phys.\ Rev.\ Lett.\ {\bf 78}, 3366 (1997).

\bibitem{Kravtsov} V.E.\ Kravtsov and B.L.\ Altshuler, 
                   Phys.\ Rev.\ Lett.\ {\bf 84}, 3394 (2000). 

\bibitem{Moha2} P.~Mohanty, 
                Ann.\ Phys.\ (Leipzig) {\bf 8}, 549 (1999);
                and cond-mat/9912263.

\bibitem{Schwab} P.\ Schwab, cond-mat/0005525.

\bibitem{Luther} A.~Luther and I.~Peschel, 
                 Phys.\ Rev.\ B {\bf 9}, 2911 (1973).

\bibitem{gs2} T.~Giamarchi and H.~Schulz,
              Phys.\ Rev.\ B {\bf 37}, 325 (1988).

\bibitem{Shankar} R.~Shankar, Int. J. Mod. Phys. B {\bf 4} 2371 (1990).

\bibitem{SetS} B.~Sutherland and B.S.~Shastry, 
                Phys.\ Rev.\ Lett.\ {\bf 65}, 1833 (1990).
 
\bibitem{GetS} T.~Giamarchi and B.S.~Shastry, 
               Phys.\ Rev.\ B {\bf 51}, 10915 (1995).

\bibitem{AetW} A.\ M\"uller-Groeling, H.A.\ Weidenm\"uller, and
               C.H.\ Lewen\-kopf,
               Europhys.\ Lett.\ {\bf 22}, 193 (1993); 
               A.\ M\"uller-Groeling and H.A.\ Weidenm\"uller,
               Phys.\ Rev.\ B {\bf 49}, 4752 (1994).

\bibitem{pang} H.~Pang, S.~Liang, and J.F.~Annett,
                Phys.\ Rev.\ Lett.\ {\bf 71}, 4377 (1993).
 
\bibitem{AetB} M.~Abraham and R.~Berkovits,
               Phys.\ Rev.\ Lett.\ {\bf 70}, 1509 (1993).

\bibitem{BPM} G.~Bouzerar, D.~Poilblanc, and G.~Montambaux,
              Phys.\ Rev.\ B {\bf 49}, 8258 (1994).

\bibitem{Kato} H.~Kato and D.~Yoshioka, Phys.\ Rev.\ B {\bf 50}, 4943 (1994).

\bibitem{tsiper} E.V.\ Tsiper and A.L.\ Efros,
                 J.\ Phys.:\ Condens.\ Matter {\bf 9}, L561 (1997);
                 Phys.\ Rev.\ B {\bf 57}, 6949 (1998).

\bibitem{peter} P.\ Schmitteckert, T.\ Schulze, C.\ Schuster, P.\ Schwab, and
                U.\ Eckern, Phys.\ Rev.\ Lett.\ {\bf 80}, 560 (1998).

\bibitem{Jeon} G.S~Jeon, S.~Wu, H.W.~Lee, and M.Y.~Choi,
                Phys.\ Rev.\ B {\bf 59}, 3033 (1999).

\bibitem{Chiappe} G.~Chiappe, J.A.~Verg\'es, and E.~Louis, Solid
                State Commun.\ {\bf 99}, 717 (1996).

\bibitem{Cheung} H.F.~Cheung, Y.~Gefen, E.K.~Riedel, and W.H.~Shih
                Phys.\ Rev.\ B {\bf 37}, 6050 (1988).

\bibitem{Okabe} T.~Okabe, J.~Phys.~Soc.~Jpn.\ {\bf 67}, 7292 (1998).

\bibitem{kohn} W.\ Kohn, Phys.\ Rev.\ {\bf 133}, A171 (1964).

\bibitem{scala} D.~Scalapino, S.R.~White and S.~Zhang,
                Phys.\ Rev.\ B {\bf 47}, 7995 (1995).

\bibitem{Leggett} A.J.~Leggett, in {\em Granular Nanoelectronics},
                  edited by D.K.~Ferry, J.R.~Barker, and C.~Jacobini,
                  NATO ASI Ser. {\bf B251} (Plenum, New York, 1991).

\bibitem{loss} D.\ Loss,
               Phys.\ Rev.\ Lett.\ {\bf 69}, 343 (1992).

\bibitem{Staff91} C.A.\ Stafford, A.J.\ Millis, and B.S.\ Shastry,
                  Phys.\ Rev.\ B {\bf 43}, 13660 (1991). 

\bibitem{Fye91} R.M.\ Fye, M.J.\ Martins, D.J.\ Scalapino, J.\ Wagner,
                and W.\ Hanke, Phys.\ Rev.\ B {\bf 44}, 6909 (1991). 

\bibitem{sjwp} P.\ Schmitteckert, R.A.\ Jalabert, D.\ Weinmann, and 
               J.-L.\ Pichard,
               Phys.\ Rev.\ Lett.\ {\bf 81}, 2308 (1998).

\bibitem{wpsj} D.\ Weinmann, J.-L.\ Pichard, P.\ Schmitteckert, and 
               R.A.\ Jalabert, 
               cond-mat/9905017.

\bibitem{PhysicaE} R.A.\ Jalabert, D.\ Weinmann, and J.-L.\ Pichard,
                Physica E, in press.

\bibitem{ulmke} M.\ Ulmke, V.\ Jani\v{s}, and D.\ Vollhardt,
                Phys.\ Rev.\ B {\bf 51}, 10411 (1995).

\bibitem{shepelyansky} D.L.~Shepelyansky, 
                       Phys.\ Rev.\ Lett.\ {\bf 73}, 2607 (1994).

\bibitem{tip2} X.~Waintal, D.~ Weinmann and J.-L.~Pichard, 
                Eur.\ Phys.\ J.\ B {\bf 7}, 451 (1999).

\bibitem{tip3} S.~De Toro Arias, X.~Waintal and J.-L.~Pichard, 
                Eur.\ Phys.\ J.\ B {\bf 10}, 149 (1999).

\bibitem{tip4} A.\ Wobst and D.\ Weinmann, 
                Eur.\ Phys.\ J.\ B {\bf 10}, 159 (1999).

\bibitem{dmrg} S.R.\ White, 
               Phys.\ Rev.\ B {\bf 48}, 10345 (1993).

\bibitem{dmrgbook} {\it Density-Matrix Renormalization -- A New
                    Numerical Method in Physics}, ed.\ by I.\ Peschel,
                    X.\ Wang, M.\ Kaulke, and K.\ Hallberg, Springer,
                    Berlin (1999).

\bibitem{epl1} X.\ Waintal,  G.\ Benenti and J.-L.\ Pichard, 
               Europhys.\ Lett.\ {\bf 49}, 466 (2000).

\bibitem{epl2} G.\ Benenti, X.\ Waintal, and J.-L.\ Pichard, 
               Europhys.\ Lett.\ {\bf 51}, 89 (2000).

\bibitem{al} A.F.\ Andreev and I.M.\ Lifshitz, 
             Sov.\ Phys.\ JETP, {\bf 29}, 1107 (1969).

\bibitem{imryma} Y.\ Imry and S.\ Ma, 
                 Phys.\ Rev.\ Lett.\ {\bf 35}, 1399 (1975). 


\bibitem{Been} C.W.J.\ Beenakker, 
               Phys.\ Rev.\ B {\bf 44} 1646 (1991).

\bibitem{benenti_gap} G.\ Benenti, X.\ Waintal, J.-L.\ Pichard, and 
                      D.L.\ Shepelyansky, 
                      Eur.\ Phys.\ J.\ B {\bf 17}, 515 (2000).

\bibitem{CandB} P.\ Cedraschi and M.\ B\"uttiker,
                J.\ Phys.\ C {\bf 10}, 3985 (1998).

\bibitem{bouch89} H.\ Bouchiat and G.\ Montambaux, 
                  J.\ Phys.\ France {\bf 50}, 2695 (1989). 

\bibitem{Staff93} C.A.\ Stafford and A.J.\ Millis,
                  Phys.\ Rev.\ B {\bf 48}, 1409 (1993). 

\bibitem{SetW} F.\ Selva and D.\ Weinmann,
               cond-mat/0003202, to appear in Eur.\ Phys.\ J.\ B (2000).

\bibitem{kambili} A.\ Kambili, C.J.\ Lambert, and J.H.\ Jefferson,
                  Phys.\ Rev.\ B {\bf 60}, 7684 (1999).

\bibitem{montambaux} F.\ Piechon and G.\ Montambaux, private communication.

\bibitem{KDS} R.\ Kotlyar and S.\ Das Sarma, cond-mat/0002304.

\bibitem{Sca} R.T.\ Scalettar, G.G.\ Batrouni, and G.T.\ Zimanyi,
              Phys.\ Rev.\ Lett.\ {\bf 66}, 3144 (1991).

\bibitem{Schr} T.\ Vojta, F.\ Epperlein, and M.\ Schreiber,
               Phys.\ Rev.\ Lett.\ {\bf 81}, 4212 (1998).

\bibitem{ilani} S.\ Ilani, A.\ Yacoby, D.\ Mahalu, and H.\ Shtrikman,
                Phys.\ Rev.\ Lett.\ {\bf 84}, 3133 (2000). 

\bibitem{expspin} D.\ Simonian, S.V.\ Kravchenko, M.P.\ Sarachik, and
                  V.M.\ Pudalov, 
                  Phys.\ Rev.\ Lett.\ {\bf 79}, 2304 (1997).

\end{thebibliography}
\end{document}